\documentclass[ip,twocolumn]{jpsj3}
\usepackage{txfonts}

\usepackage[super]{nth}
\usepackage{url}
\newcommand{\newblock}[0]{}
\newcommand{\myurl}[1]{}
\newcommand{\href}[2]{#2, \url{#1}}

\usepackage{color}
\definecolor{Blue}{rgb}{0.3,0.3,0.9}
\definecolor{Red}{rgb}{0.9,0.3,0.3}
\definecolor{Green}{rgb}{0.3,0.6,0.3}

\newcommand{\figdir}{.}

\title{Machine learning of phases and structures for model systems in physics}

\author{%
Djenabou Bayo$^{1,2}$, 
Burak \c{C}ivitcio\u{g}lu$^2$, 
Joseph J Webb$^1$, 
Andreas Honecker$^2$, and 
Rudolf A.\ R\"{o}mer$^1$\thanks{r.roemer@warwick.ac.uk}}

\inst{
$^1$Department of Physics, University of Warwick, Coventry, CV4 7AL, United Kingdom \\
$^2$Laboratoire de Physique Th\'eorique et Mod\'elisation, CNRS UMR 8089, CY Cergy Paris Universit\'e, 95302 Cergy-Pontoise, France} 

\abst{
The detection of phase transitions is a fundamental challenge in condensed matter physics, traditionally addressed through analytical methods and direct numerical simulations. In recent years, machine learning techniques have emerged as powerful tools to complement these standard approaches, offering valuable insights into phase and structure determination. Additionally, they have been shown to enhance the application of traditional methods. In this work, we review recent advancements in this area, with a focus on our contributions to phase and structure determination using supervised and unsupervised learning methods in several systems: (a) 2D site percolation, (b) the 3D Anderson model of localization, (c) the 2D $J_1$-$J_2$ Ising model, and (d) the prediction of large-angle convergent beam electron diffraction patterns.
}

\begin{document}
\maketitle

\section{Introduction}
\label{sec:introduction}

Identification of critical points separating distinct phases of matter is a central pursuit in condensed matter and statistical physics \cite{Ashcroft1976SolidPhysics,Diep2013FrustratedSystems}. This task requires a thorough understanding of the global behavior of the many-body system because phenomena may emerge that are very difficult to derive from microscopic rules \cite{Anderson1972MoreDifferent}.
Traditional analytic methods and numerical simulations have proven effective in understanding these complex systems \cite{Yu2016DimensionlessTransitions,Wang2024DistinguishingStatistics}, but they often come with limitations, particularly in high-dimensional parameter space \cite{Gomez2019AProblems}.  

Machine-learning methods, particularly supervised \cite{Alpaydin2020IntroductionEdition} and unsupervised learning techniques \cite{Hinton1999UnsupervisedComputation}, have in the last years appeared in physics as a novel strategy to bypassing some of these limitations \cite{Mehta2019APhysicists,Carleo2019MachineSciences,Schurov2024InvariantApplications}. Convolution neural networks (CNN), a class of deep, i.e., multi-layered, neural networks (DNNs) in which spatial locality of data values is retained during training, have, when coupled with a form of residual learning \cite{He2016a}, shown to allow astonishing precision when classifying images, e.g., of animals \cite{Tabak2019} and handwritten characters \cite{Zhang2017CombinationRecognition}, or when predicting numerical values, e.g., of market prices \cite{Zhao2020}. 
These \emph{supervised} learning strategies similarly yield promising predictions in identifying critical points or phases in parameter space \cite{Carrasquilla2017,ChNg2018,Tanaka2017DetectionNetworks,Huembeli2018a, Dong2019MachineTransitions, Canabarro2019UnveilingLearning, Shinjo2019MachineModel}, providing an alternative and potentially more efficient way of exploring complex systems.
By now, the evidence in favour of supervised machine-learning methods' efficacy in identifying different phases of a physical system appears compelling \cite{Carrasquilla2017,ChNg2018,Tanaka2017DetectionNetworks,Huembeli2018a, Dong2019MachineTransitions, Canabarro2019UnveilingLearning, Shinjo2019MachineModel}. 
\emph{Unsupervised} learning and semi-unsupervised learning approaches have also demonstrated the ability to reconstruct the outlines of a system's phase diagram.\cite{Wang2016DiscoveringLearning,Kottmann2020UnsupervisedDetection,Alexandrou2020TheAutoencoders,DAngelo2020LearningNetworks,Shiina2020Machine-LearningModels,Corte2021ExploringModels}. The potential to identify structural changes within a system further supports the significance of these techniques in modern scientific exploration \cite{Walker2018IdentifyingMethods}.

Among the various models studied in the context of machine learning and \emph{statistical physics}, the Ising model on the square lattice has served as an important benchmark \cite{Wang2016DiscoveringLearning,Carrasquilla2017,Morningstar2018DeepCriticality,Suchsland2018,Efthymiou2019Super-resolvingNetworks,Alexandrou2020TheAutoencoders,Walker2020DeepAutoencoder,Goel2020LearningVariables,DAngelo2020LearningNetworks,Shiina2020Machine-LearningModels,PhysRevResearch.3.013074,PhysRevResearch.3.033052,Civitcioglu2022MachineModel,Basu2022MachineSpin-shuffling,condmat8030083,PhysRevE.108.L032102,Naravane2023Semi-supervisedAutoencoders,Pavioni2024MinimalistModels} due to the simplicity of its two thermal phases, the low-temperature ferromagnet and the high-temperature paramagnet, and the ready availability of its exact solution \cite{McCoy1973TheModel} with exactly known critical temperature. 
We note that the use of ML to determine phases from just the spin configurations suggests that these themselves should contain sufficient information to identify phases, providing a level of physical insight that was, while not unknown, at least not as clear as it now seems. 
We also mention related work on multi-layer \cite{Rzadkowski2020DetectingLearning} and Potts models \cite{Fukushima2021CanModel, Giataganas2022NeuralModels, Tirelli2022UnsupervisedModel}, where the latter include the Ising model as the $q=2$ case.

Percolation can be considered as the $q\to1$ limit of the Potts model \cite{Wu1982TheModel}
and yields another class of paradigmatic models to which machine-learning techniques have been applied to identify the non-spanning and spanning phases\cite{Zhang2019MachineModels, Yu2020UnsupervisedPercolation,Shen2021SupervisedPercolation, Bayo2022MachineModel, Bayo2023TheLearning,Patwardhan2022MachineNetworks}. 
Previous ML studies have mostly used {supervised} learning in order to find the two phases via ML classification \cite{Zhang2019MachineModels,Shen2021SupervisedPercolation}. An estimate of the critical exponent of the percolation transition has also been given \cite{Zhang2019MachineModels}. The task of determining the transition threshold, $p_c$, was further used to evaluate different ML regression techniques\cite{Patwardhan2022MachineNetworks}.
For {unsupervised} and {generative} learning, less work has been done \cite{Shen2021SupervisedPercolation,Yu2020UnsupervisedPercolation,Zhang2019MachineModels}. While some successes have been reported \cite{Yu2020UnsupervisedPercolation,Cheng2021MachineModel}, other works show the complexities involved when trying to predict percolation states \cite{Shen2021SupervisedPercolation}.

Disordered electron systems provide \emph{quantum systems} with similarly rich phase diagrams. Examples are given by the Anderson insulator\cite{Li2009TopologicalInsulator}, diffusive metals\cite{Balogh2014DiffusionAlloys}, the quantum Hall\cite{Oswald2015a,Oswald2020} and quantum anomalous Hall insulators\cite{Haldane1988ModelAnomaly,Yu2010QuantizedInsulators}, Weyl semimetals\cite{Pixley2015,Meng2019,Schulz-Baldes2020}, as well as topological insulators\cite{Hasan2010,Rotenberg2011}. In these cases, the thermal states investigated for Ising-type models are replaced by quantum mechanical eigenfunctions, or variations thereof such as the local density of states (LDOS). These have specific features in each phase but, due to the random nature of these systems, precisely determining a phase from
an LDOS is difficult.\cite{Rodriguez2010,Rodriguez2011} Recent supervised learning work on the Anderson model of localization, capturing the features of eigenfunctions across the delocalization–localization transition,\cite{Ohtsuki2016DeepSystems} as well as further transfer-learning approaches to the disordered Chern insulator–Anderson insulator transition,\cite{Ohtsuki2019DrawingFunctions} have shown to allow a seemingly accurate description of phases and phase boundaries.

The power of \emph{generative} machine learning has not yet been harnessed to the same extent. This is partly because it is still a relatively novel machine learning strategy \cite{Wichert2021MachineLearning}. In brief, the difference to the supervised methods lies in the generative methods being able to seemingly {create} novel predictions which do not appear in any of the provided data. 
For example, in computer vision, generative networks construct previously non-existent high-resolution images, conditional on information from other images \cite{Isola2017Image-to-imageNetworks, Wang2018High-ResolutionGANs}.
Here, we will show how to use such generative ML strategies to study the phases for the $J_1-J_2$ Ising model, an extension of the Ising model that incorporates competing interactions across the diagonals of the Ising squares and presents a more challenging $3$-phase structure. As ML generator, we shall use a so-called variational autoencoder (VAE), a type of neural network that reconstructs a given predicted state after being trained on a selected set of states \cite{Kingma2013Auto-EncodingBayes}.

The application of ML to structure determination via electron diffraction has also blossomed in the last decade \cite{Ede2021AdvancesLearning}. ML strategies have been used to reduce the data flow in single-molecule data classification \cite{Matinyan2023MachineData}, convolutional neural nets (CNNs) were shown to help with phase reconstruction for convergent-beam electron diffraction (CBED)-based scanning transmission electron microscopy (TEM) \cite{Friedrich2023PhaseLearning} while molecular structure imaging was found to benefit from such CNNs as well \cite{Liu2021MachineStructures}. At the core of the deep learning methods employed in these works lie the same supervised DL techniques as used for phase determination. Again, generative ML for electron diffraction is not so common. Here, we will show how a so-called conditional generative adversarial network (cGAN) can be used to make accurate predictions of large-angle CBED (LACBED) images from just standard crystal information as encoded, e.g., in the usual text information\cite{Hall2006SpecificationCIF} given in the Inorganic Crystal Structure Database (ICSD) \cite{IgorLevin2018NISTICSD}, the world’s largest such database. 

\section{A brief recap of the ML approach to phases and structures}
\label{sec:ml}

\subsection{Classification and regression}
\label{sec:classregress}

Machine learning (ML) differs from traditional programming in that it does not rely on explicit rules to solve tasks. Instead, the network is expected to develop a strategy based on the input dataset to accomplish the required task.
There are three primary types of learning: supervised learning, unsupervised learning, and reinforcement learning.\cite{Wichert2021MachineLearning} Here, we will focus mainly on the first two.\cite{Mehta2019APhysicists}

Supervised learning aims to discover the optimal strategy for performing a task by using a labeled dataset.
Within supervised learning, two key tasks can be identified: classification and regression.
In classification, the ML model learns to divide data into distinct categories. Essentially, it finds an optimal representation of the dataset that separates samples into different classes.
In regression, the algorithms are trained to understand the relationship between inputs and labels, enabling them to make continuous predictions for new, unseen labels based on the given inputs. This sets regression apart from classification, as it allows the model to predict values for data not encountered during training.

The second type of learning is unsupervised learning. In this approach, the ML algorithm processes unlabeled data and is expected to uncover hidden patterns or correlations without any external guidance.
Unsupervised learning is further divided into three categories: clustering, dimensionality reduction, and association learning.
Clustering aims to group similar samples within the dataset. Dimensionality reduction seeks to simplify the data representation while retaining its essential characteristics. Association learning explores relationships between different samples in the dataset. Unsupervised learning has a wide range of applications. It can be used as a preprocessing step to reveal the structure of a dataset before supervised learning begins.\cite{Zhang2019MachineModels} It also powers generative methods, such as VAEs and GANs, which create new data samples.

\subsection{Generative ML: VAEs and cGANs}
\label{sec:generativeML}

A Variational Autoencoder (VAE) represents a relatively recent deep learning architecture that integrates standard compression techniques with the regularization strategies of machine learning, functioning simultaneously as a generative model \cite{Kingma2013Auto-EncodingBayes, Chou2019GeneratedVAE, IanGoodfellow2016DeepLearning, 
Goodfellow2014GenerativeNets}.
In essence, a VAE comprises an \emph{encoder}, which is a multilayered neural network trained on input data to generate output parameters for a variational distribution. These parameters define a low-dimensional probabilistic distribution, referred to as the \emph{latent space}.
The \emph{decoder}, another deep neural network architecture, then reconstructs the output data from the latent space, drawing samples from this space rather than selecting deterministic points.

When the latent space dimensionality, $d$, is significantly smaller than the information content of the input data, some degree of information loss is inevitable. Thus, the goal is to design the encoder and decoder in such a way that maximizes the preservation of information during encoding while minimizing the error in the reconstructed data during decoding.

To effectively train a VAE, two primary loss functions are utilized. The \emph{reconstruction loss} $\ell_{\varepsilon}$ measures the discrepancy between the input and reconstructed output during training. Additionally, the Kullback-Leibler divergence \cite{Kullback1951OnSufficiency}, which serves as a regularization term, ensures that the latent space approximates a standard normal distribution \cite{Kingma2013Auto-EncodingBayes}.
In practice, the training process involves minimizing a total loss $\ell$, which is a combination of the reconstruction loss $\ell_{\varepsilon}$ and the Kullback-Leibler loss $\ell_\text{KL}$, such that $\ell = \ell_{\varepsilon} + c  \ell_\text{KL}$, where $c$ is a hyperparameter that balances the two components \cite{Kingma2013Auto-EncodingBayes}.

GANs have emerged as a highly popular architecture for image-to-image translation tasks \cite{IanGoodfellow2016DeepLearning, Goodfellow2014GenerativeNets}. While VAEs are known to struggle with producing high-fidelity outputs, often resulting in blurriness \cite{Pathak2016ContextInpainting}, GANs inherently avoid this issue by design \cite{Isola2017Image-to-imageNetworks}. An absence of blurriness is particularly critical in quantitative electron diffraction, where clarity is essential. For this reason, we focus on conditional GANs (cGANs) \cite{Mirza2014ConditionalNets}, which are well-suited to our image-to-image task involving the learning of a mapping from an input image $x$ and random noise vector $z$ to a target image $y$, denoted as $G: {x, z} \to y$. In this context, $G$ represents the \emph{generator}.
GANs also introduce a second component, the \emph{discriminator}, denoted as $D$. The discriminator is trained to differentiate between `real' images from the dataset and `fake' images generated by $G$. This adversarial setup ensures that the generator improves over time, as the discriminator learns to recognize blurry images as fake, thereby driving the generator to produce sharper outputs.
Unlike VAEs, which rely on a predefined loss function, GANs instead learn a loss function for the desired task, solving another problem: deciding which loss function to use for comparing diffraction patterns is not apriori clear and can vary between different applications.
\renewcommand{\figdir}{percolation}

\section{Predicting percolating clusters with CNNs}
\label{sec:percolation}

This section reviews work done previously,\cite{Bayo2022MachineModel,Bayo2023TheLearning}
where we showed that standard CNNs, usually employed in image recognition ML tasks, also work very well for classifying site percolation states according to occupation probability $p$ as well as for regression when determining $p$ from such states. 
However, analyzing in detail whether spanning clusters at $p < p_c$ or non-spanning clusters at $p> p_c$ are correctly identified, we found that the same CNNs consistently fail to reflect the ground truth. Rather, it appears that the CNNs use $p$ as a proxy measure to inform their classification predictions --- a strategy that is obviously false for the percolation problem.

\subsection{The physics model of ``percolation''}

\begin{figure*}[bt]
    (a) \hspace*{-2ex}
    \raisebox{4ex}{
    \begin{minipage}[b]{0.27\textwidth}
    \includegraphics[width=0.45\columnwidth]{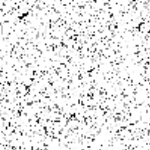}
    \includegraphics[width=0.45\columnwidth]{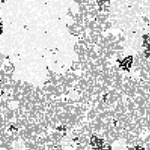}\\
    \includegraphics[width=0.45\columnwidth]{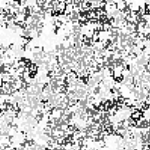}
    \includegraphics[width=0.45\columnwidth]{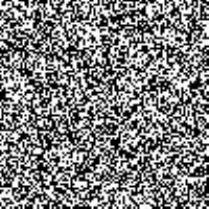}
    \end{minipage}}
    (b)
    \vspace*{-2ex}\includegraphics[width=0.3\textwidth]{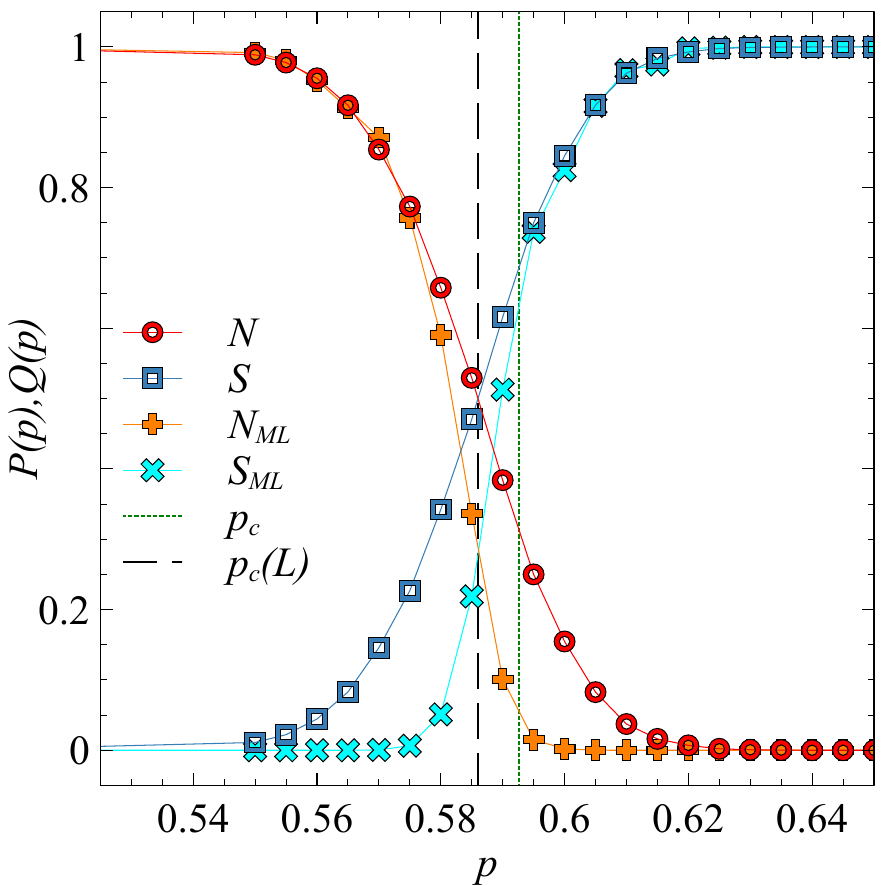}
    (c)\includegraphics[width=0.31\textwidth]{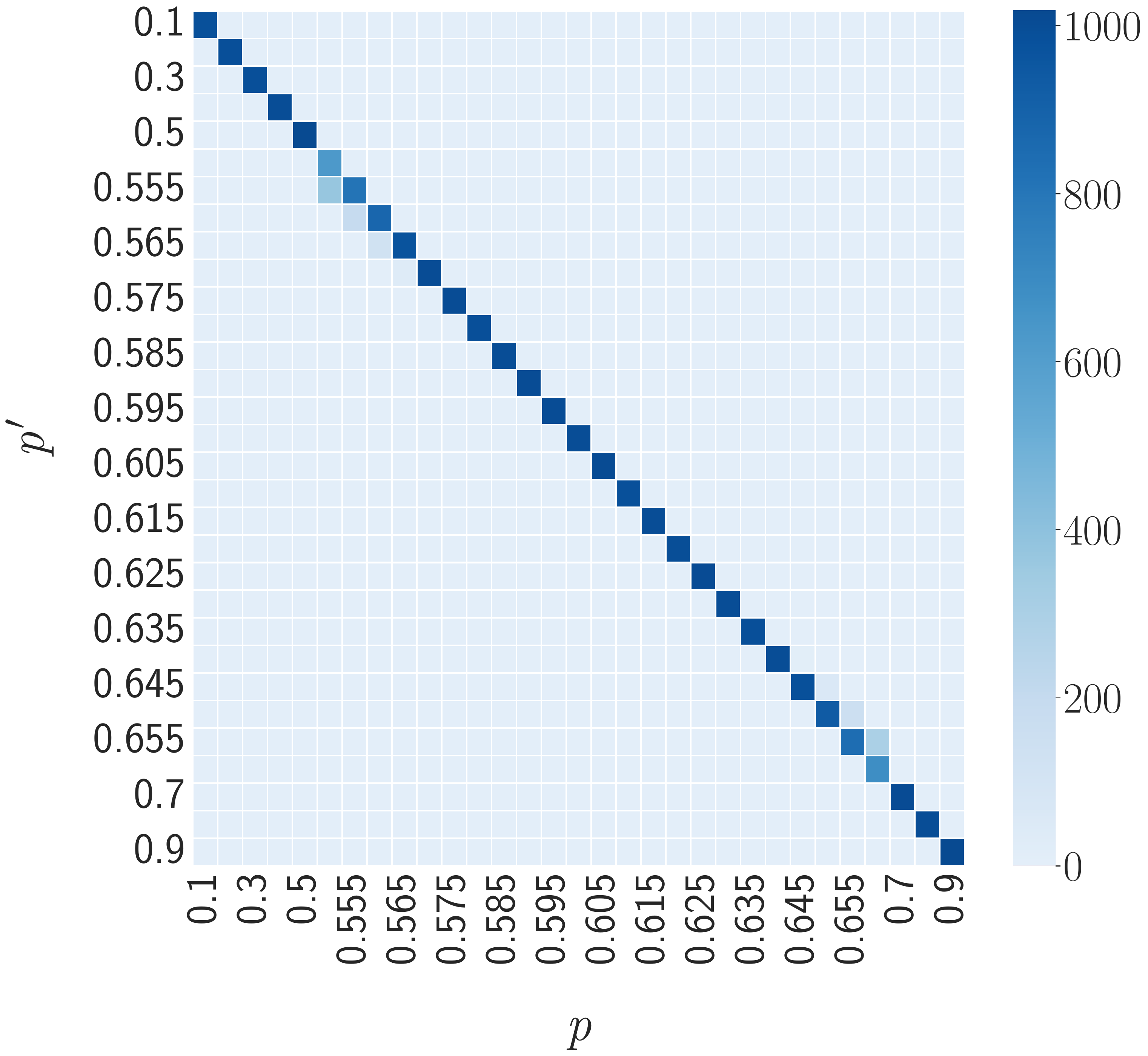} 
    \caption{
    (a) Examples of {four} percolation clusters of size $L^2=100^2$, obtained for $p=0.2<p_c$, ${p= }$ $0.6> p_c$ in the top row and $p=0.5$, i.e., {just below} $p_c$, in the bottom row. {Occupied sites are marked by small dots while empty sites are left white. Each cluster of connected sites has been identified through the Hoshen-Kopelman algorithm}. While individual clusters have been highlighted with different gray scales for the first three images, the bottom right image with $p=0.5$ shows all occupied sites in black only, irrespective of cluster identity. This latter representation is used below for the ML approach.
    (b) The blue curve (red curve) shows the probability to have a spanning, $P(p)$ (non-spanning $Q(p)$) sample in the training dataset. The cyan (orange) curve gives us the corresponding ML prediction for the probability to have a spanning (non-spanning) sample, according to the trained network. The lines connecting the symbols are only a guide to the eye. The vertical lines indicate the percolation thresholds as given in the legend.
    (c) Average confusion matrix for \textit{classification} according to $p$. The dataset used is the test data $\tau$ and the models used for predictions are those corresponding to a minimal $l_\text{c,val}$. True labels for $p$ are indicated on the horizontal axis while the predicted labels are given on the vertical axis. The color scale represents the number of samples in each matrix entry.
}
    \label{fig:percolation}
\end{figure*}

The percolation problem is well-known with a rich history across the natural sciences \cite{Broadbent1957PercolationMazes,Stauffer2018,Elliott1960EquivalenceFerromagnetism,Flory1953a,Derrida1985,Grimmett1989}. It  provides the usual statistical characteristics across a second-order transition such as, e.g., critical exponents, finite-size scaling, renormalization and universality \cite{Stauffer2018}. 
%
Briefly, on a percolation lattice of size $L \times L$, individual lattice sites $\vec{x}=(x,y)$, $x,y \in [1,L]$, are randomly occupied with \emph{occupation probability} $p$ such that the state $\psi$ of site $\vec{x}$ is $\psi(\vec{x})=1$ for occupied and $\psi(\vec{x})=0$ for unoccupied sites. 
We say that a connection between neighboring sites exists when these are side-to-side nearest-neighbors on the square lattice, while diagonal sites can never be connected. A group of these connected occupied sites is called a \emph{cluster} (cf.\ Fig.\ \ref{fig:percolation}(a)). 
Such a cluster then \emph{percolates} when it spans the whole lattice either vertically from the top of the square to the bottom or, equivalently, horizontally from the left to the right. Obviously, for $p=0$, all sites are unoccupied and no spanning cluster can exist while for $p=1$ the spanning cluster trivially extends throughout the lattice.
In Fig.\  \ref{fig:percolation}(a), we show examples of percolation clusters generated for various $p$ values.
%
%
The \emph{percolation threshold} is at $p=p_c(L)$, such that for $p< p_c(L)$ most clusters do not span while for $p > p_c(L)$
there is at least one spanning cluster. 
This can be expressed via
the quantities $P(p)$, $Q(p)=1-P(p)$ 
that denote the probabilities of the presence or absence of the spanning cluster at a given $p$, respectively (cf.\ Fig.\ \ref{fig:percolation}(b)). 
We note that $P$ is a finite-$L$ version of $\psi$ in the notation of \cite{Grimmett1989}.
We will occasionally emphasize this point using  $P_L$ and, likewise, $Q_L$.
For an infinite system ($L\rightarrow\infty$), one finds the emergence of an infinite spanning cluster at $p_{c}=0.59274605079210(2)$. This estimate has been determined numerically evermore precisely over the preceding decades \cite{Jacobsen2014} while no analytical value is yet known \cite{Grimmett1989}.

\subsection{The ML approach to the percolation problem and the generation of ML ``data''}

Several ML studies on the percolation model have been
published, mostly using {supervised} learning in order to identify the two phases via ML classification \cite{Zhang2019MachineModels,Yu2020UnsupervisedPercolation,Shen2021SupervisedPercolation,Cheng2021MachineModel,Patwardhan2022MachineNetworks}. 
%
In order to facilitate the recognition of percolation with image recognition tools of ML, we have generated finite-sized $L \times L$, with $L=100$, percolation states, denoted as $\psi_i(p)$, for the $31$ $p$-values $0.1, 0.2, \ldots$, $0.5, 0.55, 0.555, 0.556$, $\dots, 0.655, 0.66, 0.7$, $\ldots, 0.9$. For each such $p$, $N=10000$ different random $\psi_i(p)$ have been generated. 
Each state $\psi_i(p)$, $i=1, \ldots, N$, is of course just an array of numbers with $0$ denoting unoccupied and $1$ occupied sites. Nevertheless, we occasionally use for convenience the term ``image'' to denote $\psi_i(p)$.
The well-known Hoshen-Kopelman algorithm \cite{Hoshen1976a} is employed to identify and label clusters from which we (i) compute $s(p)$ and (ii) determine the presence or absence of a spanning cluster. Correlation measures have also been calculated but are not shown here for brevity.\cite{Bayo2022MachineModel,Bayo2023TheLearning}
%

We emphasize that in the construction, we took care to only construct states such that for each $p$, the number of occupied sites is exactly $N_\text{occ}= p \times L^2$ and hence $p$ can be used as exact label for the supervised learning approach. Hence $p= N_\text{occ} / L^2$ can also be called the percolation \emph{density}. 
For the ML results discussed below, it will also be important to note that the spacing between $p$ values reduces when $p$ reaches $0.5$ with the next $p$ value given by $0.55$ and then $0.555$. Similarly, the $p$ spacing increases as $0.655$, $0.66$, $0.7$. We will later see that this results in some deviations from perfect classification/regression. For reference, we now have 12 values $p=0.1, \ldots, 0.58< p_c(100)$ and 18 values $p=0.59, \ldots, 0.9> p_c(100)$. We also note that the training set contains $92.7\%$ of states without a spanning cluster below $p_c$ and $94.8\%$ are spanning above $p_c$.
We have also generated similar training and test sets for $L=200$; our results do not change significantly\cite{Bayo2023TheLearning}.
Last, all our ML results have been obtained from ten training, validation and test cycles allowing us to quote ML indicators, such as losses, accuracies, in terms of averages and their errors.\cite{Bayo2023TheLearning} Our CNN uses the \textsc{ResNet18} implementation of \textsc{PyTorch}.\cite{Paszke2019}

\subsection{Results for ML classification according to spanning or non-spanning properties}
 
The hallmark of the percolation transition is the existence of a spanning cluster which determines whether the system is percolating or not \cite{Stauffer2018}. 
%
We now want to check this and label all states according to whether they are spanning or non-spanning. From Fig.\ \ref{fig:percolation}(b), it is immediately clear that for finite-sized systems considered here, there are a non-negligible number of states which appear already spanning even when $p < p_c$ and, vice versa, are still non-spanning when $p > p_c$. Furthermore, we note that for such $L$, the difference between $p_c$ and $p_c(L)$ is large enough to be important and we hence use $p_c(L)$ as the appropriate value to distinguish the two phases.

\begin{figure}[tb]
    \centering%
    (a)\includegraphics[width=0.48\columnwidth]{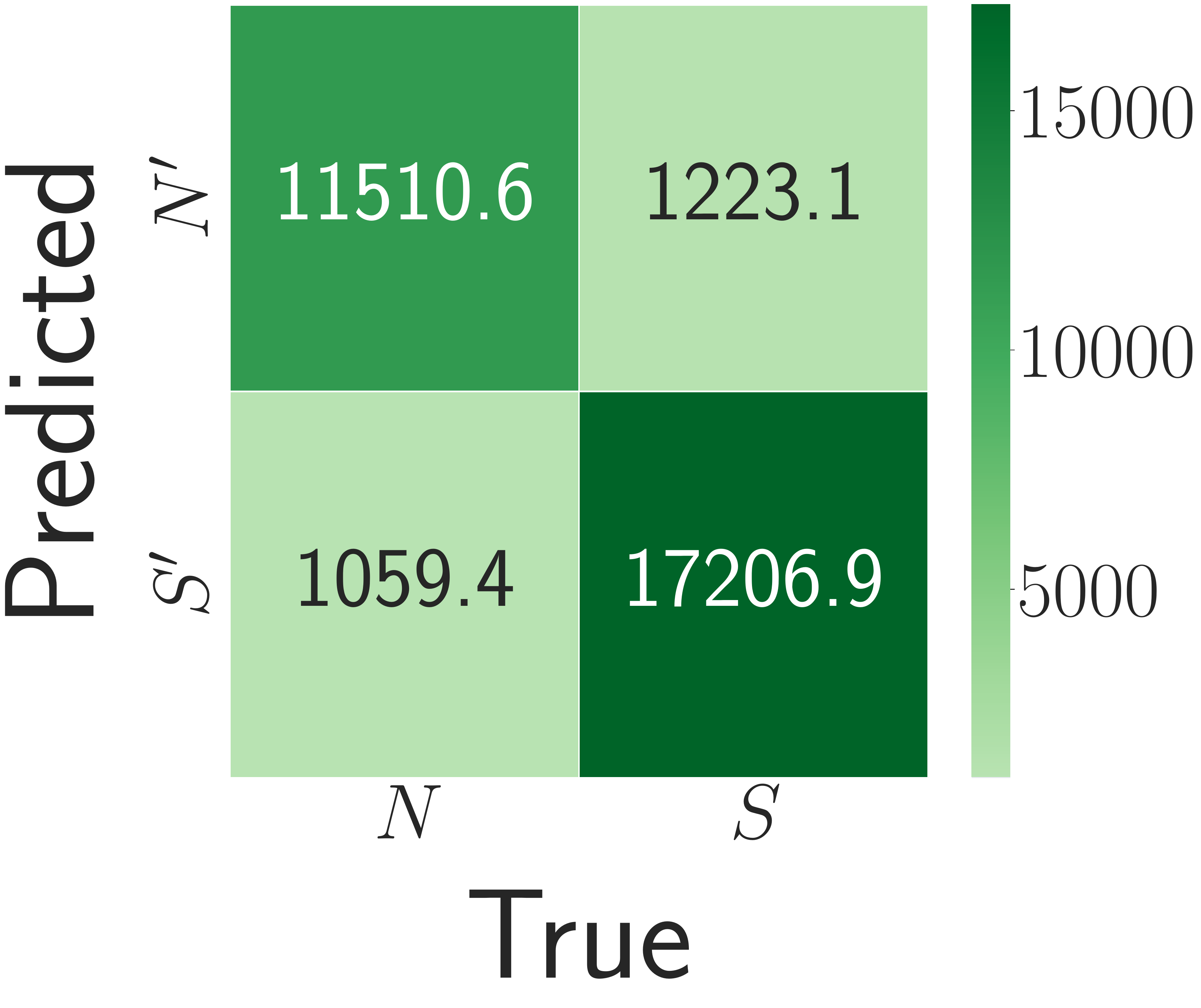}%
    \hfill%
    (b)\includegraphics[width=0.41\columnwidth]{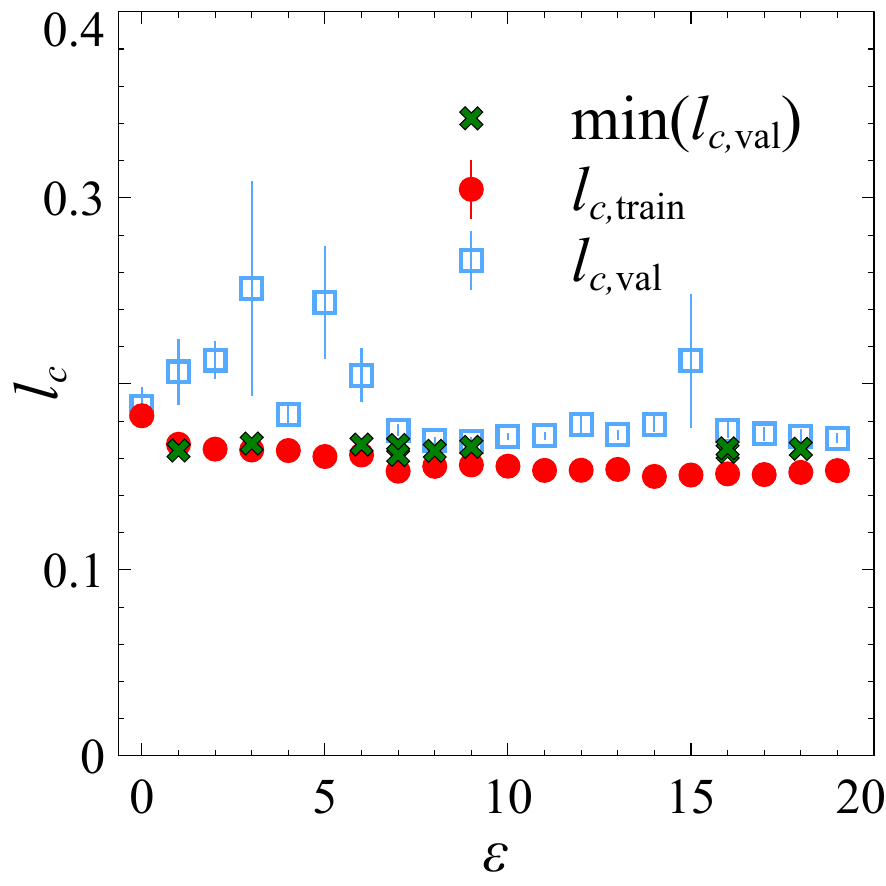}
    \caption{(a) Average confusion matrix for \textit{classification} according to  spanning/non-spanning. The dataset used is the test data $\tau$ and the models used for predictions are those corresponding to a minimal $l_\text{r,val}$. The true labels for $N$ and $S$, are indicated on the horizontal axis while the predicted labels are given on the vertical axis. 
    (b) Dependence of losses $l_\mathrm{c,train}$ and $l_\mathrm{c,val}$, averaged over ten independent training seeds, on the number of epochs $\epsilon$ for classification according to spanning/non-spanning. The circles (red solid) denote $l_\mathrm{c,train}$ while the squares (blue open) show $l_\mathrm{c,val}$. The green crosses indicate the minimal $l_\mathrm{c,val}$ for each of the ten trainings.}%
    \label{fig:ml-connectivity-total}
\end{figure}

Fig.\ \ref{fig:ml-connectivity-total} shows the averaged results after $\epsilon=20$ with a validation loss of $\text{min}_{\epsilon}[\langle l_\text{c,val} \rangle]=0.165  \pm 0.001$ (corresponding to a maximal validation accuracy $\text{max}_{\epsilon}[\langle a_\text{c,val} \rangle]= 92.702\% \pm 0.001$).
At first glance, the figure seems to indicate a great success: from the $31000$ states present in $\tau$, $11510.6$ have been correctly classified as non-spanning (i.e., $N\rightarrow N'$), and $17206.9$ as spanning ($S\rightarrow S'$) while only $1223.1$ are wrongly labeled as non-spanning ($S\rightarrow N'$) and $1059.4$ as spanning ($N\rightarrow S'$) (We note that these numbers are not integers since they are computed as averages over $10$ independent training runs \cite{Bayo2023TheLearning}). 
Overall, we would conclude that $92.6\%$ of all test states are correctly classified while $7.4\%$ are wrong. 

However, from the full percolation analysis for {$\tau$}, we {can compute that there are $11127$ states ($92.7\%$) without a spanning cluster below $p_c(L)$ while $873$ states ($7.3\%$)}
already contain a spanning cluster. {Similarly, for $p>p_c(L)$, $94.9\%$ of states, equivalent to $17075$ states, are spanning and $5.1\%$ are not, corresponding to $925$ states}. At $p_c(L)=0.585$, we furthermore have $482$ spanning and $518$ non-spanning states. Hence in total, we expect {$2280$} wrongly classified states.
Since the last number is very close to the actual number of $2282.5$ of misclassified states, this suggests that it is precisely the spanning states below $p_c(L)$ and the non-spanning ones above $p_c(L)$ which the DL network is unable to recognize. 
Let us rephrase for clarity: it seems that the CNN, when trained in whether a cluster is spanning or non-spanning, completely disregards this information in its classification outputs. 
We show that this is indeed the case by a detailed analysis of the clusters around $p_c$ as well as test sets which have been constructed to allow testing for the existence of the spanning cluster.\cite{Bayo2023TheLearning} 

In summary, when looking at $p$, classification and regression techniques for percolation states allow us to obtain good recognition with near-perfect $\langle a_\text{c,val} \rangle = 99.323\% \pm 0.003$) for classification (cf.\ also Fig.\ \ref{fig:percolation}(c)) and near-zero $\langle l_\text{r,val} \rangle =0.000062 \pm 0.000012$ average mean-square loss for regression.\cite{Bayo2022MachineModel}
On the other hand, the DL network completely ignores whether a cluster is spanning or non-spanning, essentially missing the underlying physics of the percolation problem --- it seems to still use $p$ as its main ordering measure. We believe that the root cause of the failure to identify the spanning clusters, or their absence, lies in the fundamentally \emph{local} nature of the CNN: the filter/kernels employed in the {\sc ResNet}s span a few {local} sites only. Hence it is not entirely surprising that such a CNN cannot correctly identify the essentially \emph{global} nature of spanning clusters. But it is of course exactly this global percolation that leads to the phase transition. This should serve as a warning to enthusiastic proponents of the ML approach not to ignore the physics undeservedly.

\renewcommand{\figdir}{anderson}
\section{Resolving disorder strengths from images of the 3D Anderson model}
\label{sec:localization}

\begin{figure*}[tb]
    \centering
    (a)
    \includegraphics[width=0.13\textwidth,trim=0 0 400 0, clip]{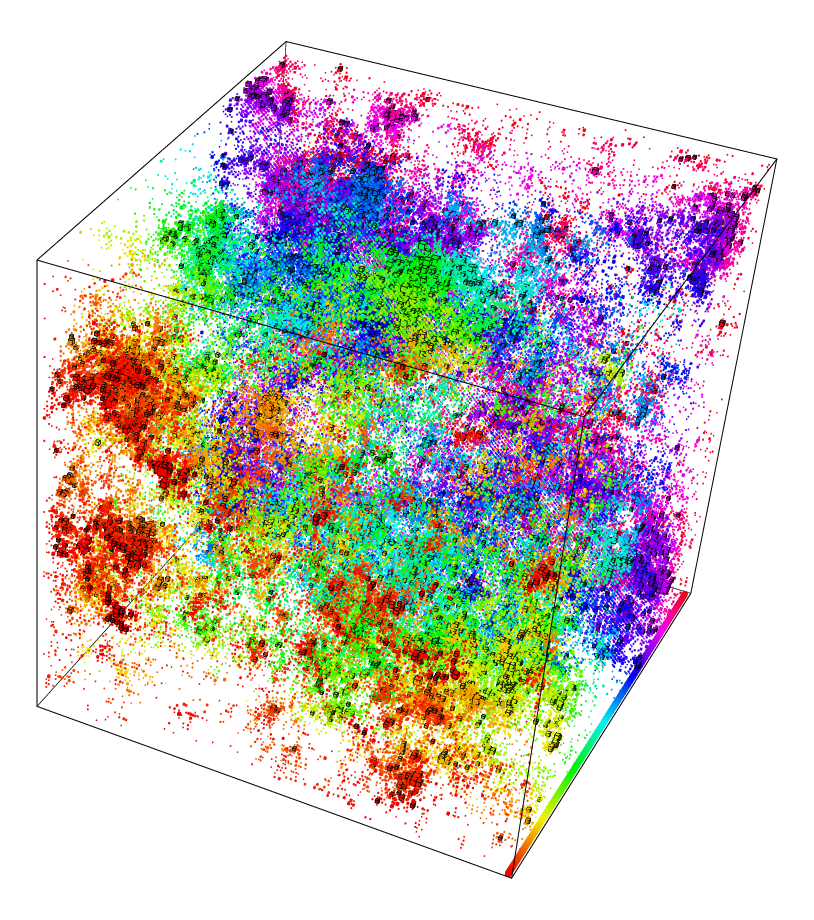}
    \includegraphics[width=0.145\textwidth,trim=250 0 0 0, clip]{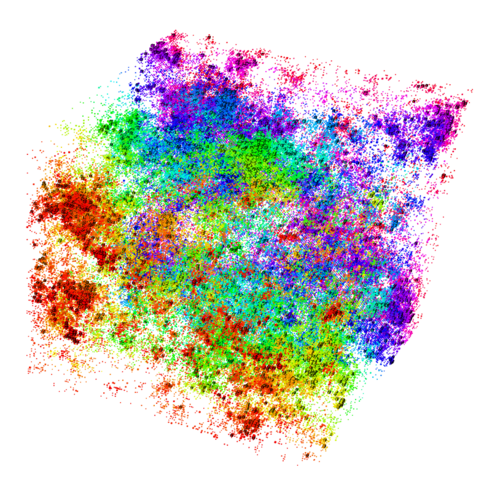}
    (b)
    \includegraphics[width=0.13\textwidth,trim=0 0 400 0, clip]{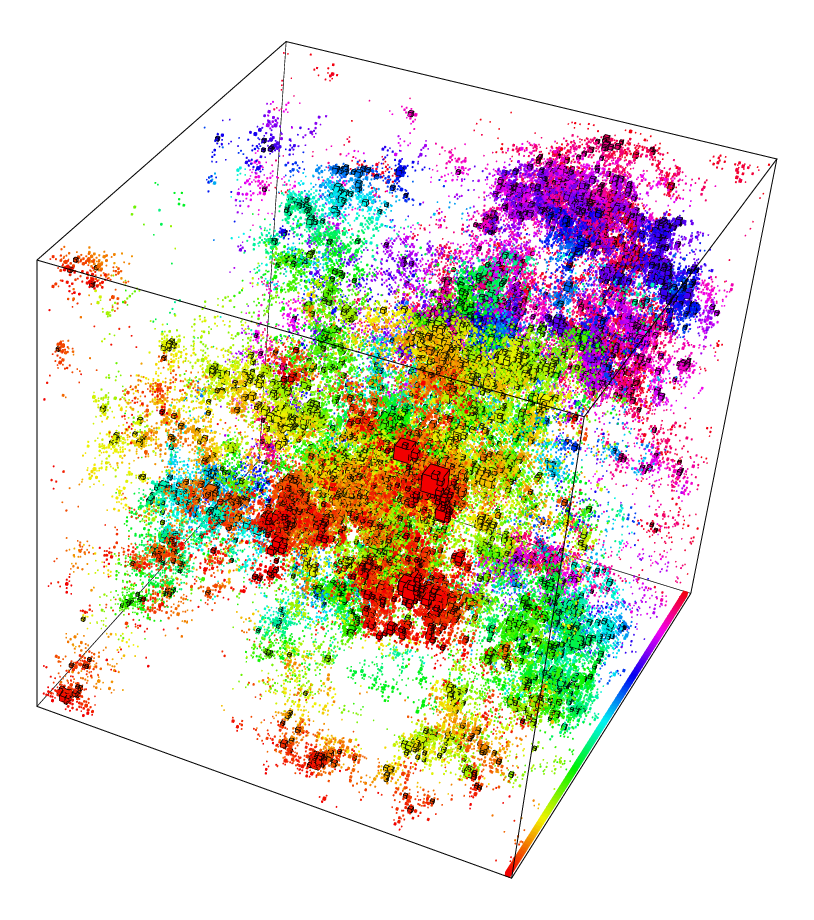}
    \includegraphics[width=0.145\textwidth,trim=250 0 0 0, clip]{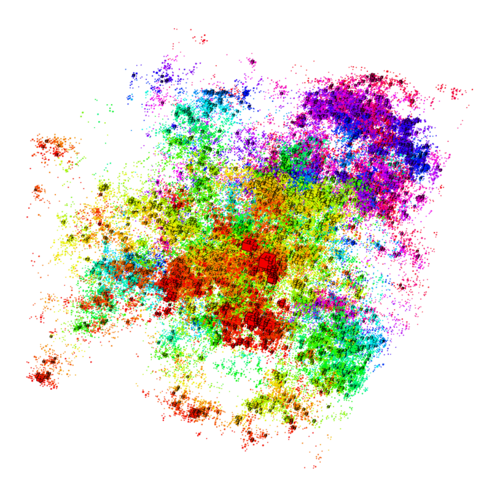}
    (c)
    \includegraphics[width=0.13\textwidth,trim=0 0 400 0, clip]{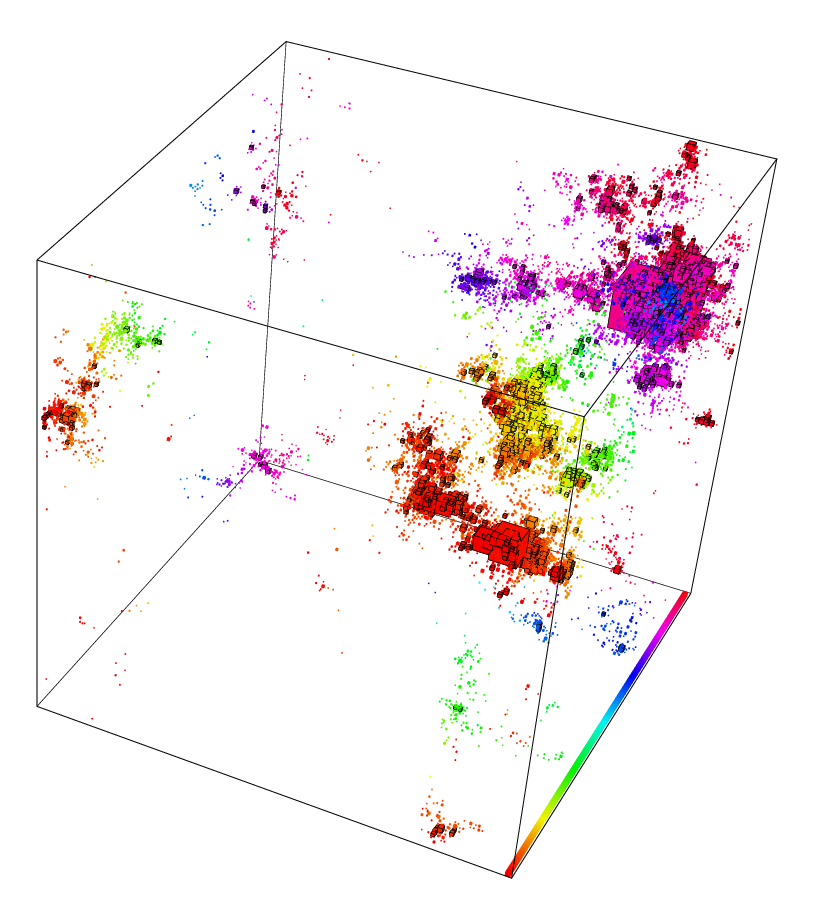}
    \includegraphics[width=0.145\textwidth,trim=250 0 0 0, clip]{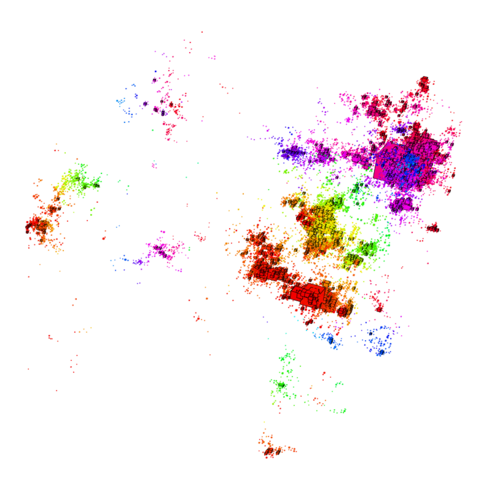}
    \caption{
    Extended (left), critical (center) and localized (right) wave function probabilities $|\psi(\vec{r})|^2$ for the 3D Anderson model with periodic boundary conditions at $E=0$ with $N =100^3$ and $w =14$, $16.5$ and $19$, respectively. 
    Every site with probability $|\psi(x,y,z)|^2$ larger than the average $1/N^3$ is shown as a box with volume $N |\psi_{E=0}(x,y,z)|^2$. Boxes with $N |\psi(x,y,z)|^2 > \sqrt{1000}$ are plotted with black edges. 
    The color scale distinguishes between different slices of the system along the axis into the page.
    In each panel, the left half is the originally constructed image while the right half shows the image in its converted PNG form with $500\times 500$ pixel resolution. Obviously, upon conversion, the black edges around the large $|\psi(x,y,z)|^2$ become less prominent and the overall black frames are also removed.
    }
    \label{fig:AMstates}
\end{figure*}

One of the hardest challenges in modern eigenvalue computation is the numerical solution of large-scale eigenvalue problems, in particular those arising from quantum physics\cite{Schenk2008b}. 
Typically, these problems require the computation of some eigenvalues and eigenvectors for systems which have up to several million unknowns due to their high spatial dimensions.
Here, the Anderson model of localization\cite{Anderson1958a} is a particularly paradigmatic model as its underlying structure involves random perturbations of matrix elements which invalidates simple preconditioning approaches based on the graph of the matrices.\cite{Elsner1999} 
Its physical importance comes from the prediction of a spatial conﬁnement of the electronic motion upon increasing the disorder -- the so-called Anderson localization.\cite{Brandes2003AndersonRamifications} When the model is used in three spatial dimensions, it exhibits a metal-insulator transition in which the disorder strength $w$ mediates a
change of transport properties from metallic behavior at small $w$ via critical behavior at the transition $w_c \sim 16.57$ to insulating behavior and strong localization at larger $w> w_c$.\cite{Evers2008} 
The 3D Anderson model hence provides us with a physically meaningful quantum problem in which to use ML strategies to distinguish its two phases, namely the metallic phase with \emph{extended} states at $w<w_c$ and the insulating phase with \emph{localized} states at $w>w_c$ (Occasionally, one might want to also study $w \approx w_c$ as a 3rd phase), while avoiding the many challenges of fully interacting quantum systems.\cite{Theveniaut2019NeuralLimitations} In this sense, it can be seen as the quantum ML test partner to complement the classical statistical physics tests available via the percolation and Ising-type models.
Similarly to the percolation model, previous ML studies have already been performed and showed good success for ML classification with CNNs to identify the two phases of the system \cite{Mano2017a,Ohtsuki2016DeepSystems,Ohtsuki2019DrawingFunctions,Cadez2023MachinePhases}. 
Here, we show that not only phases but also disorder strengths can be recovered from eigenstates of the 3D Anderson model.

\subsection{The formulation of the Anderson model in 3D}

In its usual form, the localization problem in 3D with coordinates $x, y, z$ corresponds, in the absence of a magnetic field, to a Hamilton operator in the
form of a real symmetric matrix $H$, with quantum mechanical energy levels given by
the eigenvalues ${E_n}$. The respective wave functions are simply the eigenvectors
of $H$, i.e., vectors $\psi_n(\vec{r}) \in $ for $\vec{r}=(x,y,z)$. With $N = M^3$ sites, the quantum mechanical (stationary) Schr\"{o}dinger equation is equivalent to the eigenvalue equation
$H \psi_n = E_n \psi_n$, which in site representation reads as
\begin{eqnarray}
    \lefteqn{\sum_{\sigma=\pm} \psi_n(\vec{r}+\sigma\vec{a}) + \psi_n(\vec{r}+\sigma\vec{b}) + \psi_n(\vec{r}+\sigma\vec{c})} \quad \quad \quad \quad \quad \quad \quad \quad \hfill \nonumber \\
    & = \left[ E_{n} - \varepsilon(\vec{r}) \right] \psi_n(\vec{r}),
\end{eqnarray}
with $\vec{a}=(1,0,0)$, $\vec{b}=(0,1,0)$ and $\vec{c}=(0,0,1)$ denoting the lattice vectors of a periodic, simple cubic lattice. 
The disorder usually\cite{Romer2022NumericalLocalization} enters the matrix on the diagonal, where the entries $\varepsilon_n(\vec{r})$ correspond to a spatially
varying disorder potential and are selected randomly according to a suitable distribution.\cite{Schenk2008b} Here, we shall use the standard box distribution $\varepsilon(\vec{r}) \in [-w/2,w/2]$ such that $w$ parameterizes the aforementioned disorder strength. 
For disorders $w \ll w_c$, most of the eigenvectors are \emph{extended}, i.e., $\psi_n(\vec{r})$ fluctuating from site to site, but the envelope
$|\psi_n|$ is approximately a nonzero constant. For large disorders $w > w_c$, all eigenvectors
are \emph{localized} such that the envelope $|\psi_n|$ of the $n$th eigenstate may be approximately
written as $\sim \exp \left[ -|\vec{r} - \vec{r}_n|/\xi(w) \right]$ with $\xi(w)$ denoting the \emph{localization length} of the eigenstate. 
Directly at $w = w_c$, the last extended states at $E=0$ vanish. 
The wave function vector $\psi_{E=0}(\vec{r})$ appears simultaneously extended and localized and has multifractal properties \cite{Rodriguez2010,Rodriguez2011}. 
In Fig.\ \ref{fig:AMstates}, we show examples of such states. 

In order to numerically distinguish the two (or three) phases mentioned before, one usually needs to (i) go to rather large system sizes of order $N^3=10^6$ to $10^8$ and (ii) average over many different realizations of the disorder, i.e., compute eigenvalues or eigenvectors for many matrices with different diagonals.\cite{Brandes2003AndersonRamifications,Evers2008,Rodriguez2010,Rodriguez2011}
In the present work, we concentrate on the computation of a few eigenvalues and corresponding eigenvectors for the physically most interesting case around the critical disorder $w_c$ and in the center of the spectrum $\sigma(H)$, i.e., at $E = 0$, for large system sizes.

\begin{figure*}[t]
    \centering%
    (a)\includegraphics[width=0.315\textwidth]{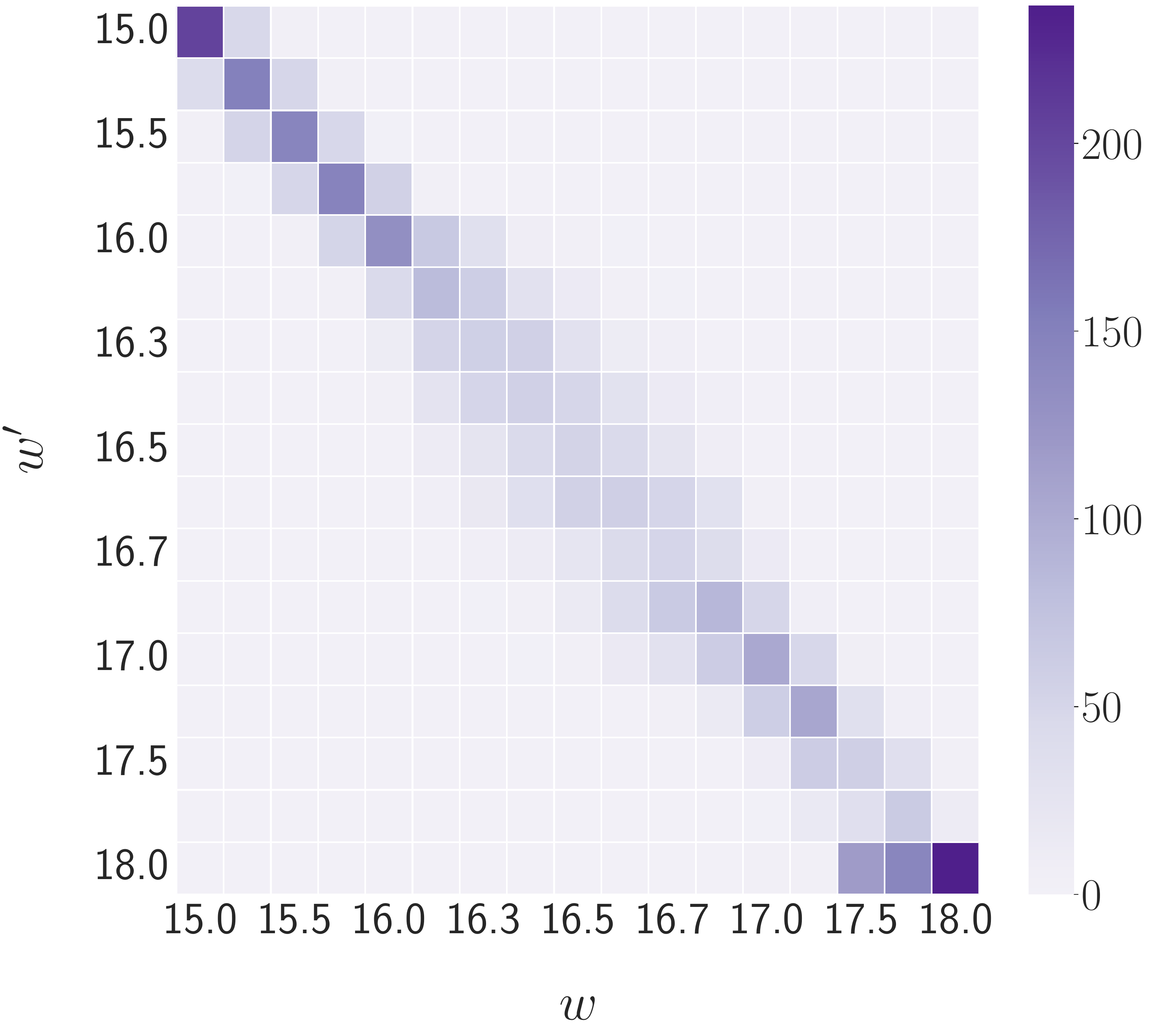}
    (b)\includegraphics[width=0.30\textwidth]{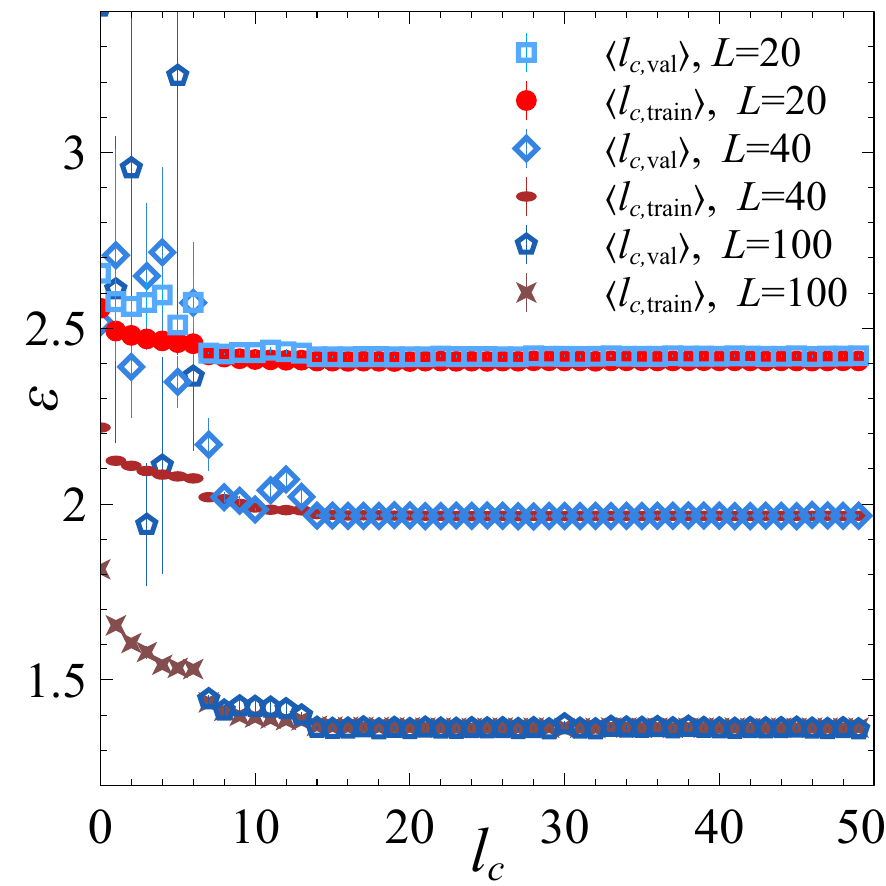}
    (c)\includegraphics[width=0.30\textwidth]{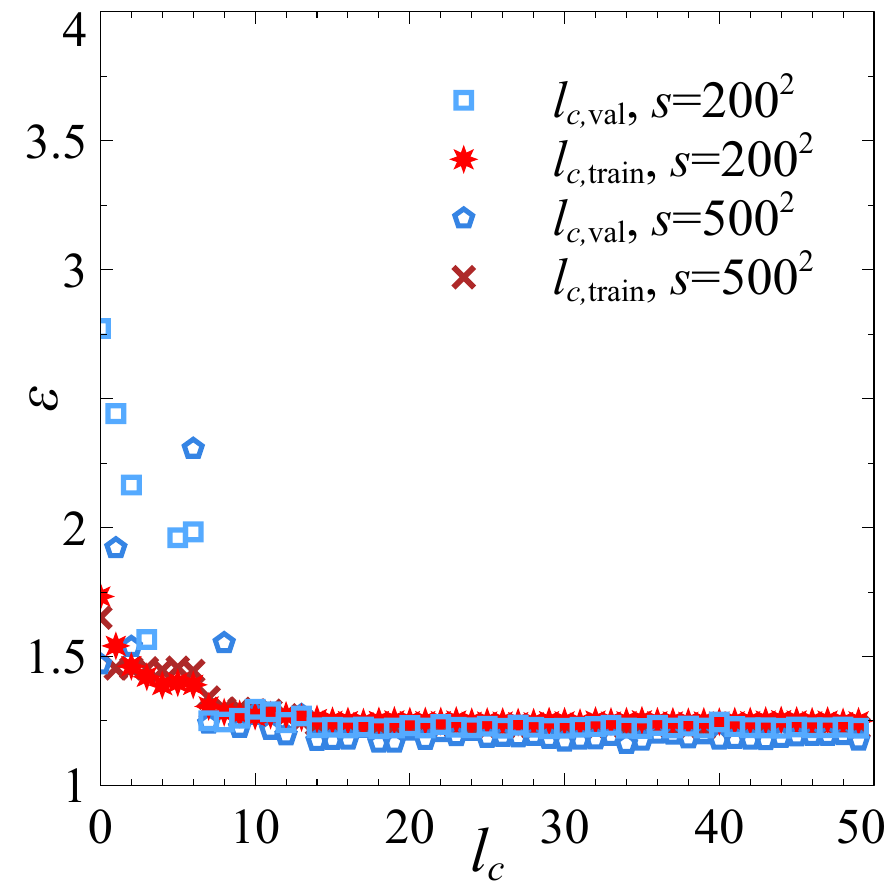}
    \caption{ 
    (a) Average confusion matrix for \emph{image} classification of the $17$ disorders $w= 15$, $15.25$, $\ldots$, $16$, $16.2$, $\ldots$, $17$, $17.25$, $\ldots$ $18$ for system size $N=100$ and image resolution $100\times 100$. The dataset used is the test data $\tau$ and the models used for predictions are those corresponding with a minimal $l_\text{c,val}$.
    (b) Dependence of losses $l_\mathrm{c,train}$ and $l_\mathrm{c,val}$ on the number of epochs $\epsilon$ for disorder classification and the three system sizes $N=20^3, 40^3$ and $100^3$ (a). The squares (blue open) denote $l_\mathrm{c,train}$ while the circles (red solid) show $l_\mathrm{c,val}$.
    (c) Epoch dependence of $l_\mathrm{c,train}$ and $l_\mathrm{c,val}$ for $N=100^3$ and two image resolutions $200\times 200$, and $500\times 500$. Symbols are as in (b), $s$ indicates the size of the images.}
    \label{fig:image-class-disorders}
\end{figure*}

\subsection{Hamiltonian eigenfunctions as data}

The square-normalized eigenstates $\psi_n= \sum_{x,y,z} \psi_n(x,y,z) |x,y,z\rangle$ have been numerically obtained using the {\sc Jadamilu} library \cite{Bollhofer2007JADAMILU:Matrices}. The $|x,y,z\rangle$ indicate the orthonormal Wannier basis in the usual tight-binding formulation.
For the $17$ disorders $w = 15, 15.25, \ldots, 16, 16.2, \ldots, 17, 17.25, \ldots 18$ we consider for training and validation a previously used dataset\cite{Rodriguez2010,Rodriguez2011} with $5000$ disorder realization for each disorder and system sizes $N= 20^3, 30^3, \ldots, 100^3$. 
For all the data,\cite{Rodriguez2010,Rodriguez2011} we have considered a single eigenstate per sample (disorder realization) with energy close to $E = 0$. 
This is costly in terms of computing time but essential to avoid the noticable correlations that exist between eigenstates of the same sample.\cite{Rodriguez2010,Rodriguez2011} 
In addition, we have generated, for each of the disorders, $500$ independent test wave functions at $E=0$, i.e., using random numbers with different seeds. 


In order to be able to use standard 2D image recognition machine learning tools, we represent the $\psi_n$ graphically as in Fig.\ \ref{fig:AMstates}. We remove the black box and the color scale before using the images for training, validation and testing purposes. Furthermore, the images are converted from their original postscript, using the {\sc ImageMagick} set of routines, and rendered as portable network graphics ({\sc PNG}) in the pixel resolutions of $s=100\times 100$, $200\times 200$ and $500\times 500$. This conversion results in some changes in the visual presentation as shown in Fig.\ \ref{fig:AMstates}.

\subsection{ML models and results}


Previous ML studies of the Anderson model use CNNs composed of $6$ convolutional layers and a fully connected layer to identify the extended and localized phases from the $|\psi(x,y,z)|^2$ \cite{Ohtsuki2016DeepSystems,Ohtsuki2019DrawingFunctions,Cadez2023MachinePhases}.
Here, our goal is to expand on these results show that a \textsc{ResNet18}, as used in section \ref{sec:percolation}, can also recover the value of $w$ used in images made from these $|\psi|^2$.
%


We first establish the capacity of the \textsc{ResNet18} to identify the two phases of the 3D Anderson model of localisation from images (not shown here). Here, we want to train a network to identify individual disorder values. Following a similar strategy as in section \ref{sec:percolation}, we train our network for $17$ disorder values $w= 15, 15.25, \ldots, 16, 16.2, \ldots, 17, 17.25, \ldots 18$ for fixed $N=20^3$, $40^3$, $100^3$ and $s=100^2$. 
After training the $17$ disorder values for $N=20^3$, we obtain a $\min_{\epsilon}[\langle l_\text{c,val} \rangle]=2.408 \pm 0.003$ (corresponding to an accuracy of $\max_{\epsilon}[\langle a_\text{c,val} \rangle] =15.9\% \pm 0.2$). 
At first, the performance of the network on this system appears to be rather limited. From the confusion matrix obtained after training (not shown), we notice that only the smallest and largest disorders, i.e., $w=15$ and $w=18$, are perfectly classified. 
We increase the size of the system and train our network for $N=40^3$. Following the training we reach $\min_{\epsilon}[\langle l_\text{c,val} \rangle]=1.951 \pm 0.004$ (corresponding to an accuracy of $\max_{\epsilon}[\langle a_\text{c,val} \rangle] =25.7\% \pm 0.2$). Looking at the metrics in Fig.\ \ref{fig:image-class-disorders} (b), we notice the decrease of $\langle l_\text{c,val} \rangle$ and $\langle l_\text{c,val} \rangle$ between the training for $N=20^3$ and $N=40^3$. 
Still, the apparent improvement in the performance of the network is not yet convincing. 
We finally train for $N=100^3$ and $s=100 \times 100$. We obtain $\min_{\epsilon}[\langle l_\text{c,val} \rangle]=1.327 \pm 0.006$ (corresponding to an accuracy of $\max_{\epsilon}[\langle a_\text{c,val} \rangle] =43.3\% \pm 0.3$). 
This is an increase of almost $18\%$. Even though the accuracy is still less than $50\%$, the network seems to be getting better at recognizing the $w$ values. The confusion matrix obtained after this training is given in Fig.\ \ref{fig:image-class-disorders}(a). Clearly, the matrix is heavily diagonally dominant: misclassifications appear to exist mostly between directly adjacent disorder values. Thus, while the training does not result in a perfect recognition of $w$'s, it is nevertheless already very good in recognizing the vicinity of each $w$, even very close to the metal-insulator transition.
%
%
Increasing the size of the input images to $s=200 \times 200$ does not help to provide significant improvement. After training we obtain  $\min_{\epsilon}[\langle l_\text{c,val} \rangle]=1.216 \pm 0.003$ (corresponding to an accuracy of $\max_{\epsilon}[\langle a_\text{c,val} \rangle] =47.9\% \pm 0.2$). Furthermore, training for such a large input leads to a substantial increase in training time.

In summary, we find that even using images of eigenstates allows to distinguish the phase of the 3D Anderson model well, while the classification of $w$ values proceeds with nearly the same accuracy as in the case of classifying $p$ for percolation in section \ref{sec:percolation}. Furthermore, increasing the system size from $N=20^3$ to $100^3$ improves the predictions considerably. Such finite-size effects remind us rather reassuringly that the ML strategies are obviously subject to the same physics constraints as standard approaches.

\renewcommand{\figdir}{j1j2}
\section{Predicting phases of the $J_1$-$J_2$ Ising model with VAEs}
\label{sec:j1j2}

The $J_1$-$J_2$ Ising model serves as a still relatively simple system to illustrate an already more complex $3$-phase behavior.\cite{Swendsen1979MonteN2,Landau1980PhaseInteractions,Binder1980PhaseInteractions,Landau1985PhaseCouplings,Moran-Lopez1993First-orderInteractions,
Malakis2006MonteInteractions, Monroe2007PhaseModel, dosAnjos2008PhaseLattice, Kalz2008PhaseInteractions,Watanabe2023Non-monotonicSystems,Yoshiyama2023Higher-orderLattice,Li2024ALattice}.
With $J_1$ denoting the nearest-neighbor interaction, the competing second-neighbor interaction $J_2$ gradually suppresses the ordering temperature, until it vanishes completely when $J_2={|J_1|}/2$ \cite{Swendsen1979MonteN2,Landau1980PhaseInteractions}. Furthermore, beyond this point, a new ordered ``superantiferromagnetic phase'' appears.
The universality class of the transition into the superantiferromagnetic phase has been investigated early on  
\cite{Swendsen1979MonteN2,Landau1980PhaseInteractions,Binder1980PhaseInteractions}, but still continues to attract attention since its nature remains controversial.\cite{Watanabe2023Non-monotonicSystems,Yoshiyama2023Higher-orderLattice,Li2024ALattice}
There is at least also one investigation of this model on the D-wave quantum annealer \cite{Park2022FrustratedMachine}
and a small number of machine-learning investigations
\cite{Corte2021ExploringModels,Basu2022MachineSpin-shuffling}.

\begin{figure*}[tb]
     \begin{center}
    \raisebox{0\columnwidth}{(a) }\includegraphics[width=0.25\columnwidth]{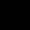}\hfil%
    \raisebox{0\columnwidth}{(b) }\includegraphics[width=0.25\columnwidth]{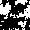}\hfil%
    \raisebox{0\columnwidth}{(c) }\includegraphics[width=0.25\columnwidth]{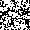}\hfil%
    \raisebox{0\columnwidth}{(d) }\includegraphics[width=0.25\columnwidth]{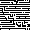}\hfil%
    \raisebox{0\columnwidth}{(e) }\includegraphics[width=0.25\columnwidth]{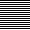}
    \end{center}
    \begin{center}
    \raisebox{0\columnwidth}{(f) }\includegraphics[width=0.25\columnwidth]{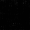}\hfil%
    \raisebox{0\columnwidth}{(g) }\includegraphics[width=0.25\columnwidth]{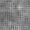}\hfil%
    \raisebox{0\columnwidth}{(h) }\includegraphics[width=0.25\columnwidth]{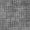}\hfil%
    \raisebox{0\columnwidth}{(i) }\includegraphics[width=0.25\columnwidth]{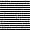}\hfil%
    \raisebox{0\columnwidth}{(k) }\includegraphics[width=0.25\columnwidth]{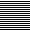}
    \end{center}
  \caption{(a-e) Five illustrative spin configurations of the $J_1$-$J_2$ Ising model on a periodic $30 \times 30$ square lattice.
  (f-k) Predicted spin configurations from (a-e), respectively, using our VAE learning.
  %
  Panels (a, b, c) keep $J_2 = 0.1$ fixed while increasing $T$ from 
  (a) ferromagnetic at $T=0.1$ to 
  (b) $T=1.975$ near the ferromagnetic-to-paramagnetic transition and 
  (c) a configuration deep in the paramagnetic phase at $T=4.0$.
  Panels (d, e) have $J_2=0.8$ and then decrease $T$ from 
  (d) $T=1.575$ near the paramagnetic-to-superantiferromagnetic transition to
  (e) a superantiferromagnetic configuration at $T=0.1$. 
  In (f-k), the parameters are as in (a-e).
  In all cases, the black squares correspond to up spins while white is for down spins. In (f-k), the values {in the interval $[-1, 1]$ are denoted by the gray squares} in the panels. }
    \label{fig:sample-configurations}
\end{figure*}
\subsection{Definition of the $J_1$-$J_2$ Ising model}

The Hamiltonian of the $J_1$-$J_2$ Ising model can be expressed as 
\begin{equation}
H_{J_1J_2} = 
-J_1 \,\sum_{\langle i,j \rangle} s_i\, s_j + 
J_2 \,\sum_{\langle\langle i,j \rangle\rangle} s_i \,s_j\, ,
\label{eq:hamiltonian}
\end{equation}
where $s_i$ represents the spin at site $i$, which can be either up ($+1$) or down ($-1)$; $\langle i,j \rangle$ refers to nearest-neighbor pairs, $\langle\langle i,j \rangle\rangle$ denotes next-nearest neighbor pairs, while $J_1$, $J_2 \geq 0$  signify the interaction strengths between the nearest and next-nearest neighbors, respectively.
Our chosen sign conventions in Eq.~(\ref{eq:hamiltonian}) lead to a ferromagnetic coupling for $J_1$ pairs while next-nearest neighbors prefer to align in an antiferromagnetic structure.\cite{Civitcioglu2024PhaseLearning}
The three distinct phases of the model correspond to (i) a low-temperature, low-$J_2$ \emph{ferromagnet}, (ii) a low-temperature, high-$J_2$ \emph{superantiferromagnet} and (iii) the high-temperature \emph{paramagnet}. We illustrate spin configurations representative of these phases and close-to-phase transitions in Fig.~\ref{fig:sample-configurations}.
In this work, we review recent work aiming to predict the three phases with a generative VAE, using a spin-adapted mean-squared error $\varepsilon$ as ML cost function.\cite{Civitcioglu2024PhaseLearning} 

\subsection{Generating states as ML training data via the Metropolis Monte-Carlo approach}
\label{sec:j1j2-data}

To generate the necessary input data for the training of the VAE, we utilize the Metropolis algorithm, a well-established method for simulating statistical models at finite temperature \cite{Metropolis1953EquationMachines,Newman1999MontePhysics,Berg2004MarkovAnalysis,Landau2014APhysics}.
In the present investigation, we initially focus on a system size of $30\times30$ with periodic boundary conditions \cite{Corte2021ExploringModels,Acevedo2021PhaseLearning}. In order to assess the influence of the size of the system, we also investigate  $60\times60$ and $120\times120$ square lattices.
Equilibration of the model can be difficult, in particular in the regime of
$J_2 \approx \left|J_1\right|/2$ \cite{Kalz2008PhaseInteractions}. We assure proper thermalization by successively cooling our configurations for fixed $J_2/\left|J_1\right|$.\cite{Civitcioglu2024PhaseLearning} 
We set the energy scale with $J_1=1$.

For $J_2=0$, we are back to the nearest-neighbor Ising model with known critical temperature $T_{c,\text{Ising}} \approx 2.269$ \cite{Kramers1941}. We can therefore confidently start our exploration of the as-yet unknown phase diagram by choosing an initial temperature range of $0 \leq T \leq 4 \approx 2 \times T_{c,\text{Ising}}$ containing $T_{c,\text{Ising}}$. 
We also know that the ferromagnetic-to-superantiferromagnetic transition is at ${J_2}=1/2$.\cite{Kalz2008PhaseInteractions} Hence we choose a range for ${J_2}$ from $0$ to $1.5$. Should we later see that these ranges do not suffice to capture all phases, we could further increase the maximal $T$ and $J_2$ values.
%
Using $\Delta T = 0.025$, we thus proceed with a set ${\cal T}$ of $|\mathcal{T}|=157$ temperatures with $T \in [0.1,4]$ for $T \in {\cal T}$. The Monte-Carlo construction is repeated with different random numbers until we have $C=40$ configurations for each temperature at the given values of $J_2$. Let ${\cal J}_2 = \{
0$, $0.1$, $0.2$, $0.3$, $0.4$, $0.45$, $0.48$, $0.49$, $0.495$, $0.5$, $0.505$,
$0.51$, $0.52$, $0.55$, $0.6$, $0.65$, $0.7$, $0.8$, $0.9$, $1$, $1.2$, $1.5 
\}$ denote the $|\mathcal{J}_2|=22$ chosen distinct values. 
In total, this results in a dataset containing $|\mathcal{T}|\times |\mathcal{J}_2| \times C = 157 \times 22 \times 40 = 138\,160$ independent configurations for a given system size. 
%
%


\subsection{Reconstruction of the phase diagram using single-region VAEs}
\label{sec:j1j2-results-singleVAE}

\begin{figure*}[tb]
     \begin{center}
{(a)}\includegraphics[width=0.95\columnwidth]{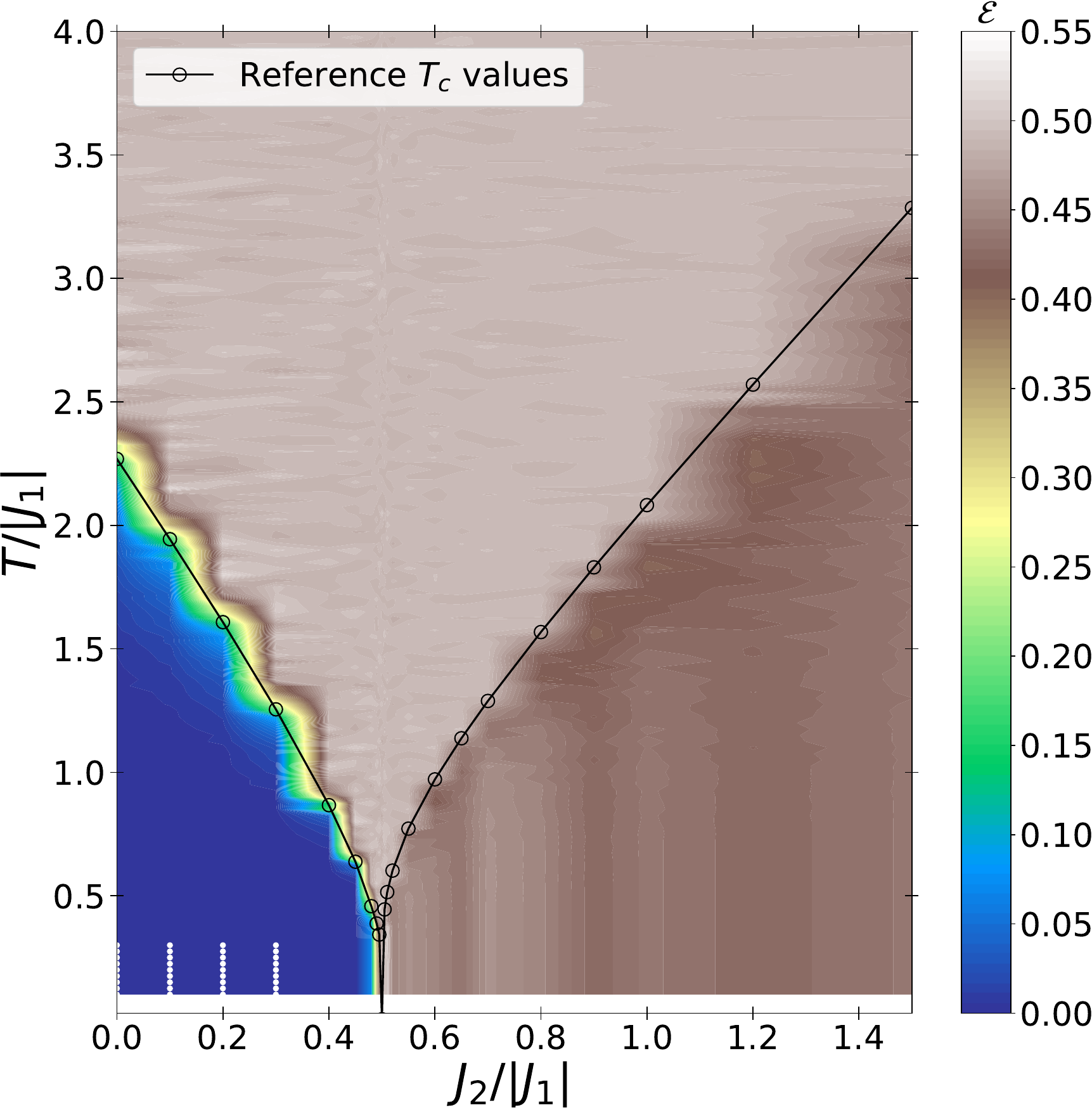} 
{(b)}\includegraphics[width=0.95\columnwidth]{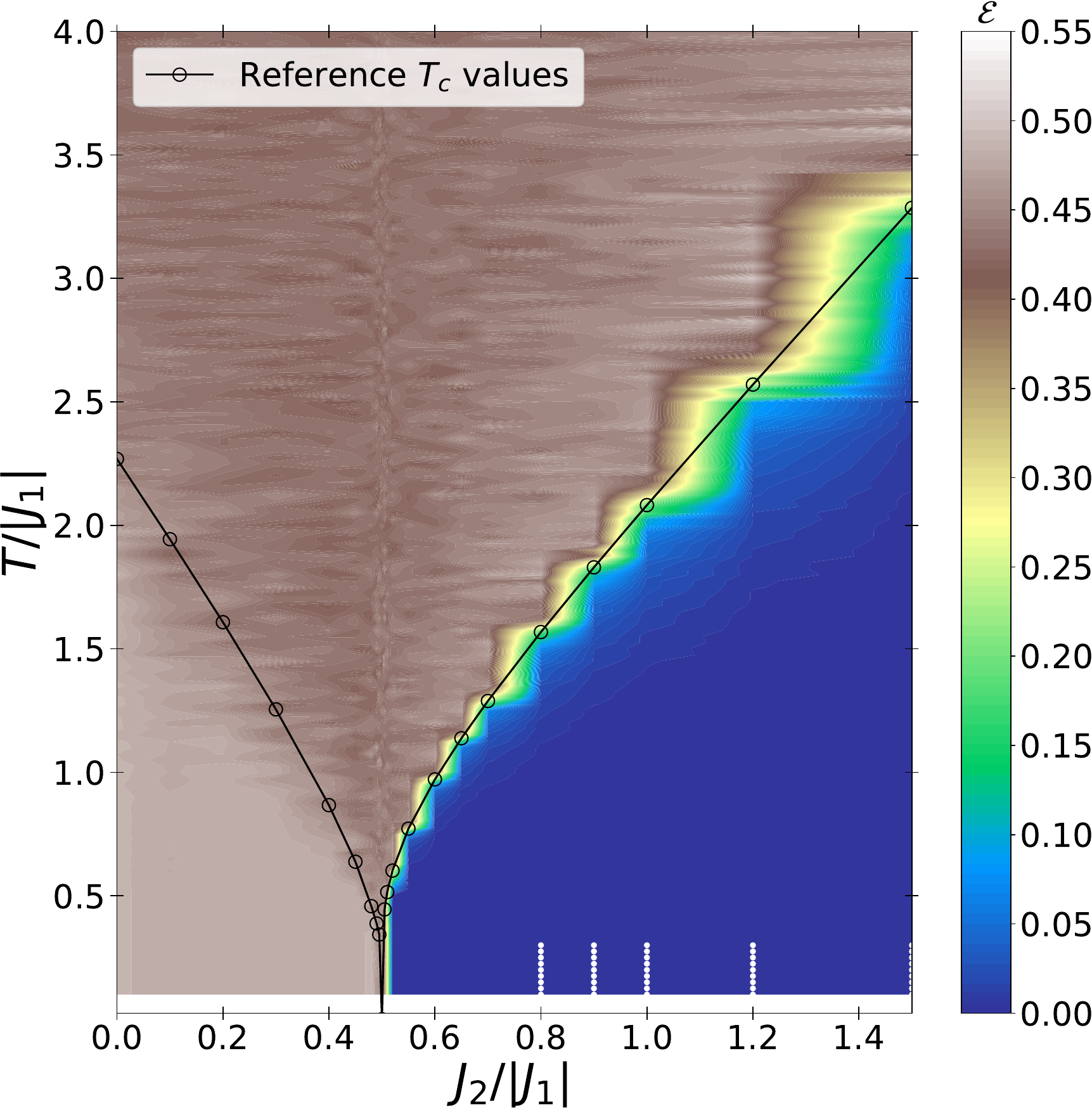}
    \end{center}
  \caption{
  Minimal averaged error $\varepsilon$ for the VAE-based reconstruction of the $J_1$-$J_2$ model's phase diagram. The results correspond to $L=30$.
  Panel 
  (a) represents the in-phase learning from the low-$J_2$ region $\rho_{\text{low-}J_2}$ and 
  (b) gives results for the in-phase learning from the high-$J_2$ region $\rho_{\text{high-}J_2}$. 
  The $(T,J_2)$ data points of various training regions are indicated by small white dots for each $(T,J_2)$ pair (usually these are closely spaced and hence appear as vertical lines).
  In all panels, $\circ$ symbols connected by black lines denote known reference phase boundaries.\cite{Kalz2008PhaseInteractions}.
  }
\label{fig:results-VAE-phase-diagram}
\end{figure*}

We can now use the VAE architecture to identify the phases of the $J_1$-$J_2$ model as a function of $T$ and $J_2$ for constant $J_1=1$. Details of the VAE implementation can be found elsewhere.\cite{Civitcioglu2024PhaseLearning}
%
We start the training of the VAE for $T \ll T_{c,\text{Ising}}$ in two distinct regions, namely (i) $J_2 < 1/2$ and (ii) $J_2 > 1/2$. Consequently, we have two restricted training data regions
$\rho_{\text{low-}J_2}$ and $\rho_{\text{high-}J_2}$. In order to have a reasonable amount of training data, we use all $40$ values for each $(T, J_2)$ in each training region. For the results underlying  Fig.~\ref{fig:results-VAE-phase-diagram}(a), this amounts to $1440$ training configurations in $\rho_{\text{low-}J_2}$, while for  Fig.~\ref{fig:results-VAE-phase-diagram}(b), we have $1800$ configurations in $\rho_{\text{high-}J_2}$. 

From  Fig.~\ref{fig:results-VAE-phase-diagram} we see that indeed two distinct regions emerge. The low-$T$, low-$J_2$ region shown in (a) is clearly separated from the rest of the $(T, J_2)$ plane. Similarly, panel (b) establishes a low-$T$, high-$J_2$ region.
We note that in both cases, the $\varepsilon$ values in the low/high-$J_2$ regions are close to zero, while in the other regions we have $\varepsilon\approx 0.5$.
This value suggests that in both cases, the out-of-region configurations have about $50\%$ of spins different, in agreement with the behavior in the known phases.
We can therefore conclude the existence of two low-$T$ regions identified in Fig.~\ref{fig:results-VAE-phase-diagram}. By exclusion, the third region corresponds to $\varepsilon \approx 0.5$ from both trainings.

Indeed, these regions agree very well with the previously established phase boundaries shown in Fig.~\ref{fig:results-VAE-phase-diagram}. The $\varepsilon$ values of $0$, $0.25$, and $0.5$ indicate best, random, and worst reconstruction possible, respectively, compatible with the spin configurations in each phase. 
Clearly, the regions with $\varepsilon \approx 0$ correspond to the ordered ferro- and superantiferromagnetic phases in Fig.~\ref{fig:results-VAE-phase-diagram} (a) and (b), respectively.
Further results with similar ML strategies as well as on a direct comparison of states can be found elsewhere.\cite{Civitcioglu2024PhaseLearning}

\renewcommand{\figdir}{felix}
\section{Microscopy with GANs}
\label{sec:microscopy}

Convergent-beam electron diffraction (CBED) \cite{Spence1992ElectronMicrodiffraction,Tanaka1994Convergent-beamDiffraction} is a transmission electron microscopy (TEM) technique with unparallel sensitivity \cite{Beanland2013DigitalPicture}. Its origins date back nearly $100$ years to pioneering work\cite{Kossel1939ElektroneninterferenzenBundel} and its modern applications include crystal symmetry classification \cite{Buxton1976ThePatterns, Tanaka1983Point-groupDiffraction, Tanaka1983Space-groupDiffraction}, lattice parameter determination \cite{Saunders1999QuantitativeMeasurements, Zuo1998AMatching, Kaiser1999ApplicationSubstrates}, strain \& defect analysis \cite{Armigliato2003ApplicationDevices, Kramer2000AnalysisCBED, Cherns1989ConvergentMultilayers, Morniroli1996AnalysisDiffraction}, and more \cite{Midgley1996QuantitativeBonds}. 
However, CBED sees the majority of its use in symmetry determination \cite{Buxton1976ThePatterns} and charge density refinement \cite{Zuo1999DirectCu2O} and is still lacking in popularity when compared to the more established structure solution and refinement methods of X-ray and neutron diffraction  \cite{Beanland2021RefinementDiffraction, Hubert2019StructurePatterns}.
Collecting the necessary amount of high-quality diffraction data from a TEM, to construct a large-angle CBED (LACBED) image, is one of the inherent challenges of the method. Here, modern computer-controlled TEM setups offer a clear advantage and can make the task near automatic \cite{Beanland2013DigitalPicture}.
A perhaps even more constraining challenge lies in the fact that the complexity introduced by multiple scattering of electrons as they propagate through the specimen \cite{Spence1992ElectronMicrodiffraction} requires sophisticated modelling techniques to construct the theoretical predictions to compare with TEM results.
To make CBED a more accessible approach, there have been two major computational methods developed: (i) the Bloch-wave method \cite{Zuo1995OnMethod, Spence1992ElectronMicrodiffraction, 2018ElectronMicroscopy, Hubert2019StructurePatterns}, and (ii) Multislice \cite{Cowley1957TheApproach,  VanDyck1984TheMicroscopy, Chuvilin2005OnSimulation, Kaiser2006ProspectsCalculation, Kirkland2010AdvancedMicroscopy, 2018ElectronMicroscopy}. Whilst both have seen success in accurately generating CBED patterns, they even today remain computationally resource- and time-intensive, often well beyond what a standard desktop computer can provide \cite{BeanlandFelix:Software}.

\subsection{Selection of training data}
\label{sec:icsd}
\label{sec:data-training}

For our aim to generate LACBED patterns via ML, we require a large body of data in which crystal structure information has been paired with corresponding bright field LACBED images. Experience from previous such machine learning tasks in computer vision \cite{Wichert2021MachineLearning,2023Kaggle:Science} and related applications \cite{Ohtsuki2016DeepSystems,Ohtsuki2019DrawingFunctions,Acevedo2021PhaseLearning,Bayo2022MachineModel,Bayo2023TheLearning}, as well as in the previous sections, suggests that often more than $10,000$ such training pairs are needed. On the scale necessary for a successful model, it is infeasible to use experimental data for such patterns. 
Fortunately, the Inorganic Crystal Structure Database (ICSD) \cite{IgorLevin2018NISTICSD}, as the world’s largest such database, provides ready access to the full structural information for more than $240,000$ crystals in the form of a `Crystallographic Information File' (CIF), a standard text file format \cite{Hall2006SpecificationCIF}.

\begin{figure*}[tb]
\centering
        \includegraphics[scale=1]{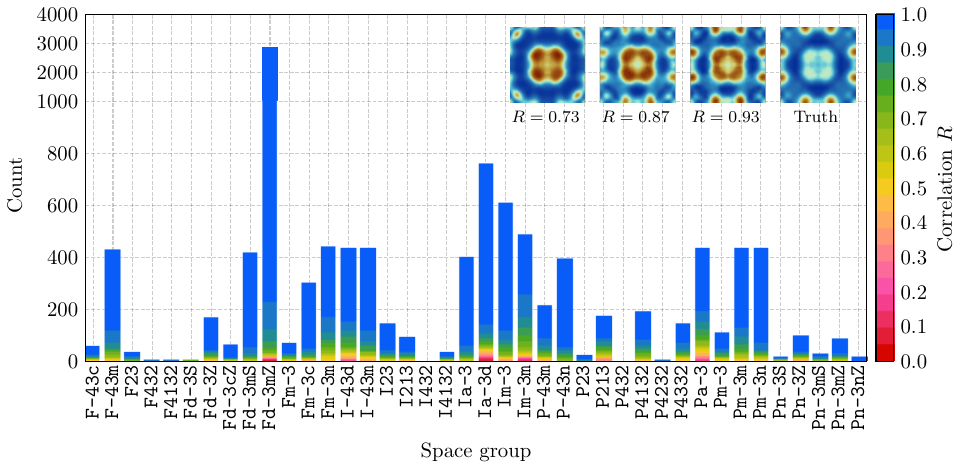}

    \caption{%
    The distribution of crystal structure data entries in each of the $36$ ($+4$ from alternative origin choices) cubic space groups as obtained from the ICSD. The color scale denotes the cross-correlation index $R$, see \eqref{eq:zmcc}, obtained for each structure in the indicated space group using the trained cGAN, sorted from overall lowest (green) to overall highest (red) loss. We note that the vertical scale above $1000$ has been compressed for clarity.
    The LACBED images shown on the top correspond to different $R$ values to give a more intuitive interpretation.
    }
    \label{fig:space_groups}
    \label{fig:pred-LACBED-space_groups}
\end{figure*}

Direct training with structured textual data, as available in the CIFs, is still a major challenge for machine learning tasks \cite{Halevy2009TheData,Spasic2020ClinicalReview}. Even small changes in, e.g., numerical values of the lattice parameters, can have major changes in the resulting CBED patterns. 
On the other hand, existing image-to-image translation tasks have been primarily optimised for 2D data. To harness this knowledge, we need a feasible way to represent the CIF information in 2D image form as well. 
Fortunately, the projected electron potential $\rho$ is a convenient such image representation. Since we have decided to concern ourselves only with cubic (isometric) crystals, we can nicely project the electron potential along $z$ to a 2D image as shown in Fig.\ \ref{fig:input_data_examples} below.
Using the structure factors $F(\mathbf{g})$ of the crystal, obtained from \textsc{Felix} \cite{BeanlandFelix:Software}, we generate the projected potential using 
$%
\rho (\mathbf{r}) \propto 
\sum_{g} F (\mathbf{g}) \cdot \exp[-2\pi i \mathbf{g} \cdot \mathbf{r}]$
.   
Here, the $\mathbf{g}$ are the lattice vectors of the unit cell in reciprocal space. 
We normalize the resulting image of electron potential strength and also restrict their size to $128\times 128$. 
Note that in requiring all of these inputs to be the same image dimensions for the machine learning model, we lose information regarding the size of the crystal.

Next, we need a method to construct the LACBED patterns corresponding to each CIF and projected electron potential $\rho$. We employ \textsc{Felix}, an open-source software implementation of the Bloch-wave method\cite{2018ElectronMicroscopy,Zuo1995OnMethod} for generating LACBED images originally developed in part by two of us \cite{BeanlandFelix:Software}.
\textsc{Felix} has been shown to provide atomic coordinate refinements with sub-picometer accuracy \cite{Beanland2021RefinementDiffraction,Hubert2019StructurePatterns}, and can accurately simulate LACBED patterns\cite{Beanland2013DigitalPicture}.
The software takes as input a CIF, beam parameters, microscope settings, crystal settings, and the desired beam direction. Most of these values were calibrated previously\cite{Beanland2021RefinementDiffraction}. 
In our simulations, we only consider the $(0,0,0)$ beam direction for simplicity. The other simulation parameters used here are provided in the code accompanying the present work \cite{Webb2024GitHubWephy/ai-diffraction}. We use \textsc{Felix} to generate LACBED images of size $128\times 128$. Such sizes are sufficient for many computer-vision-based machine learning tasks \cite{Wichert2021MachineLearning}, whilst remaining small enough to allow the generation of results on a large scale for our dataset.

Our strategy in generating the necessary input from the information provided in the ICSD is then as follows: (i) We convert the textual information provided by each crystal's CIF into the normalized projected electronic potential, i.e., a 2D image. (ii) we compute, via the Bloch-wave code \textsc{Felix}, the corresponding bright-field LACBED images.
While the construction of the electron potential images is very fast, generating the LACBED dataset takes a few weeks using bespoke high-performance computing architecture. 
In the current proof-of-principle work, we focus on the $12454$ CIFs each corresponding to a unique \emph{cubic} crystal.
We ultimately have a dataset of $12454$ image pairs of size $128\times 128$, with pixel values between $0$ and $255$. Each pair consists of a crystal's projected electron potential and its simulated $(0,0,0)$ LACBED diffraction pattern. This dataset is publicly hosted \cite{Webb2023FelixPatterns}.

Whilst we use as many cubic crystals as we can, since after all, a much higher number is still desired, we encounter significantly imbalanced data in many areas. For example, when resolved according to their space group classification, we find that the ICSD data is highly imbalanced. As shown in Fig.\ \ref{fig:space_groups}, some space groups contain less than $10$ ICSD entries while others have many thousands. 
It is well known that machine learning methods, and in particular our chosen adversarial network architecture, suffer in their predictive strength when using imbalanced data \cite{Saini2023TacklingReview}. Hence it could be worthwhile in future studies to include other crystal data from the ICSD beyond the cubic ones.

\subsection{Results}

\begin{figure}[tb]
    \centering
    intensity\\
    \includegraphics[width=0.95\columnwidth]{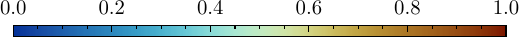}\\[2pt]
    mean prediction \\[1pt]
    \includegraphics[width=0.45\columnwidth]{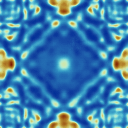}\hspace{3pt}
    \includegraphics[width=0.45\columnwidth]{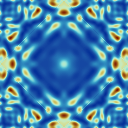}\\
    median prediction \\[1pt]
    \includegraphics[width=0.45\columnwidth]{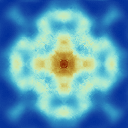}\hspace{3pt}
    \includegraphics[width=0.45\columnwidth]{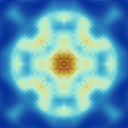}\\
    \nth{67} percentile prediction \\[1pt]
    \includegraphics[width=0.45\columnwidth]{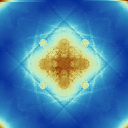}
    \hspace{3pt}
    \includegraphics[width=0.45\columnwidth]{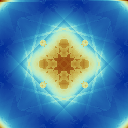}\\
    \quad \hfill (a) prediction \hfill (b) \textsc{Felix} simulation \hfill \quad
    \caption{Three LACBED patterns from unseen crystal structures.
    The left column shows the ML LACBED pattern predicted for the [001] projected electron potential of a unit cell, and the right column its (ground truth) {\textsc{Felix}} Bloch-wave simulation.
    From top to bottom, the materials (and ICSD codes) of the three crystals are are USnNi$_4~\texttt{(54390)}$, $\beta-$Mn~$\texttt{(163414)}$ and Sr$_8$Ga$_{16}$Ge$_{30}~\texttt{(153755)}$. Their correlations, given by Eq.~\eqref{eq:zmcc}, are $R = 0.934\pm0.011, 0.990\pm0.002$, and $0.997\pm0.001)$ respectively.
    Furthermore, from top to bottom, the preditions are based on taking different statistical pixel-to-pixel measures, as indicated in the respective headings, for the losses to find the optimal choice.
    Computation of each simulated image via \textsc{Felix} takes about $400$ seconds in 48 CPU cores whilst the ML result is generated in under 20 milliseconds.
    All images are $128\times 128$ pixels, have normalized amplitudes and use the same color scale as indicated at the top of the image.}%
    \label{fig:input_data_examples}
\end{figure}

We train a cGAN to create LACBED patterns by providing the projected electron density as input.\cite{Webb2024GitHubWephy/ai-diffraction} In Fig.\ \ref{fig:input_data_examples} we show some results.
Before going into details on how we create these images, we start by noting that the ground truth LACBED images shown in the figure, with whom we compare our predictions, needed about $400$ seconds each to be constructed by the Bloch-wave method on a high-performance compute cluster while our predicted LACBED images arrived within $20$ milliseconds on a modern, i.e., GPU-supported, desktop. 

The images given in Fig.\ \ref{fig:input_data_examples} show, in the left column, the computed project potential for three crystal structures from different space groups, namely \texttt{F$\bar{4}$3m}, \texttt{P$4_1$32} and \texttt{Pm$\bar{3}$n} from top to bottom. Each such projected potential has been normalized. The comparison between the CBED predictions of the cGAN and the expected behavior from the \textsc{Felix} results show an overall good agreement, in particular w.r.t.\ the underlying two-fold symmetries. While we use the standard mean-squared error in training the cGAN, in Fig.\ \ref{fig:pred-LACBED-space_groups}, we report a \emph{shifted} zero-mean normalized cross-correlation ﬁt index $R$ for pixel intensities, 
\begin{equation}
    R(\mathbf{y}, \hat{\mathbf{y}}) = \frac{1}{2} + \frac{1}{2 n^2}\sum_{i,j}^{n} 
    \frac{y_{ij}-\langle \mathbf{y} \rangle}{\sigma(\mathbf{y})} \cdot
    \frac{\hat{y}_{ij}-\langle \hat{\mathbf{y}} \rangle}{\sigma(\hat{\mathbf{y}})} ,
    \label{eq:zmcc}
\end{equation}
which has often been used in CBED image comparison studies \cite{Beanland2013DigitalPicture}. 
In the normalization used here, the value $R=1$ corresponds to a perfect fit, while $R=0$ is perfectly anti-correlated. The values $R=0.5$ and $0.75$ emerge when $1/2$ or $3/4$ of the image pixels are correlated and $1/2$ and $1/4$ are anti-correlated, respectively. Also, $R=0.5$ corresponds to two images with uncorrelated intensities. 
Our resulting $R$ values for the LACBED image reconstruction, as given in the caption of Fig.\ \ref{fig:input_data_examples}, indicate an overall very good agreement.

\renewcommand{\figdir}{.}
\section{Conclusions and Outlook}
\label{sec:conclusions}

The learning aspects of DL networks are often referred to as ``black boxes'', highlighting that it appears occasionally surprising how a DNN arrives at its classification, regression or generative predictions. On the other hand, it is exactly this lack of apriori imposed basic descriptors that allows a DL architecture to variationally construct its own set of descriptors to achieve an optimal prediction.
%
So when ML succeeds in classifying states of Ising-type, percolation, and Anderson models, this also shows that the phase information must be encapsulated directly in the states alone, even for those relatively close to the phase boundaries as shown by the overall good reconstruction of these phases. While this was not unknown before or unexpected,\cite{Edwards1975TheoryGlasses,Evers2008,Rodriguez2010,Rodriguez2011} it is nevertheless an interesting qualitative insight to have re-emphasized. 
Conversely, this also suggests that simply comparing states with each other, by mean-squared deviations, $R$ correlation or otherwise, might also be an alternative quantitative method for phase diagram construction - as already demonstrated.\cite{Civitcioglu2024PhaseLearning}

The caveat discovered when studying the globally spanning cluster for the percolation problem with locally focused CNNs, i.e., the failure of such CNNs to correctly identify the percolating cluster,\cite{Bayo2023TheLearning} furthermore suggests that even the power of modern ML approaches can fail when the underlying physics is ignored.\cite{Theveniaut2019NeuralLimitations}
In this context it is also important to mention that the cGAN predictions for the outcome of electron interference experiments, i.e., the LACBED intensities, do not somehow circumvent the quantum mechanical measurement problem. Rather, they simply provide a good interpolation to the various diffraction solutions of the electron dynamical scattering problem provided by the Bloch wave calculations the cGAN was trained on. 

Last, the review given here clearly reflects the prejudices and preferences of its authors in selecting the applications of ML to physics. Many other applications and application areas have been ignored such as Boltzmann machines\cite{Mehta2019APhysicists} and the extremely interesting approaches to finding states of many-body systems\cite{Carleo2016,Zhu2023HubbardNet:Networks,Pfau2024AccurateNetworks}.

\begin{acknowledgment}
\paragraph{Acknowledgments} J.J.W.\ and R.A.R.\ gratefully acknowledge discussions with Richard Beanland, Warwick, on the results presented here for the LACBED reconstructions. This work forms part of an ongoing project of all three of us.
D.B., A.H., and R.A.R.\ are grateful for co-tutelle funding via the EUtopia Alliance. R.A.R.\ also very grateful acknowledges the CY Initiative of Excellence (grant ``Investissements d’Avenir'' ANR-16-IDEX-0008).
The computations presented here used the University of Warwick's Research Technology Platform (RTP Scientific Computing) and the Sulis Tier 2 HPC platform hosted by the RTP. Sulis is funded by EPSRC Grant EP/T022108/1 and the HPC Midlands+ consortium. 
\end{acknowledgment}





\profile{Djena Bayo}{graduated from the Sorbonne Universit\'{e} in 2017. Following this, she obtained a Master's degree from CY Cergy Paris Universit\'{e} in 2019. As part of the Master's program, she did an internship under the supervision of Vita Ilakovac on ``Calculation of vibrations in core excited molecules" at the Laboratoire de Chimie Physique-Mati\`{e}re et Rayonnement (Sorbonne Universit\'{e}). In  2019, she joined R\"{o}mer at the University of Warwick as a co-tutelle PhD student with Honecker at CY Cergy Paris Universit\'{e}. She defended her PhD thesis\cite{PhDBayo24} in April 2024.}

\profile{Burak \c{C}ivitcio\u{g}lu}{was born in Antalya, T\"{u}rkiye. After a first degree at  Ko\c{c} University / Istanbul, he obtained his Master degrees from the Universit\'{e} Paris Diderot and CY Cergy Paris Universit\'{e}. In 2020, he started a PhD there with Honecker, informally co-supervised by R\"{o}mer. He is about to defend his PhD thesis.\cite{PhDCivitcioglu24}}

\profile{Joe Webb}{
was born in Essex in the UK.
He recently received a BSc in Mathematics and Physics from the University of Warwick.
There he founded and published two editions of the Poincar\'{e} student magazine and collaborated with R\"{o}mer on ML applications to LACBED patterns.
He is currently a scholar at Worcester College, Oxford,
studying for an MSc, with plans to pursue a PhD.
}

\profile{Andreas Honecker}{was born in T\"ubingen, Germany in 1967 and obtained first his diploma and then in 1995 his PhD degree in physics from the University of Bonn under the supervision of Werner Nahm and G\"unter von Gehlen, respectively. He went on his post-doctoral journey to the FU Berlin, SISSA Trieste, and ETH Z\"urich. After that, he worked at the TU Braunschweig where he obtained his Habilitation in 2003 under the direction of Wolfram Brenig. 
He continued on to the Georg-August-Universit\"at G\"ottingen, with partial support via a Heisenberg fellowship,
where he was awarded a professor title (apl.). He held a number of further fixed-term professorial appointments at Hannover, Strasbourg, and Lyon before he became full professor at CY Cergy Paris Universit\'e, France in 2014. Beyond broad scientific interests in statistical physics and condensed matter theory, he also contributes to administrative obligations. In particular, he served as director of the physics department and is presently serving as adjoint director of the Institut des Sciences et Techniques of CY Cergy Paris Universit\'e. He proudly supports new publishing initiatives such as SciPost Physics as editor.}

\profile{Rudo R\"{o}mer}{was born in Gedern, Hessen, Germany in 1966. He attended the Wolfgang-Ernst Gymnasium in B\"{u}dingen, 
studied with Robert Schrader for a Dipl.-Phys.\ at the FU Berlin, obtained a PhD with Bill Sutherland at the University of Utah in 1994, postdoc'ed as Feodor-Lynen fellow with Sriram Shastry at the IISc in Bangalore, Dieter Vollhardt at the RWTH Aachen, and achieved his Habilitation qualification with Michael Schreiber at the TU Chemnitz in 2000. In 2002 he was appointed at the University of Warwick, UK, where he heads the disordered quantum systems research group. Since then he has spent research stays at the MPIPKS in Dresden, the ICCMP at UN Brasilia (now in Natal), the Chinese Academy of Sciences at the Wuhan Institute of Physics and Mathematics, was appointed Lotus (Fu Rong) Visiting Professor at Xiangtan University, Xiangtan, Hunan province, China, and as Senior Fellow at CY Cergy Paris Universit\'{e}. He is on a number of editorial and advisory boards and editor-in-chief for Physica E.}







\end{document}